\documentclass[acmsmall,nonacm,authorversion,screen]{acmart}

\AtBeginDocument{%
  \providecommand\BibTeX{{%
    \normalfont B\kern-0.5em{\scshape i\kern-0.25em b}\kern-0.8em\TeX}}}

\usepackage{multirow, makecell}
\usepackage{siunitx}
\usepackage{xcolor,colortbl}
\usepackage{subfig}

\usepackage{adjustbox}

\newcommand{\added}[1]{#1}
\newcommand{\deleted}[1]{}
\newcommand{\itemdeleted}[1]{}

\usepackage{acmart-taps}

\begin{document}

\title[SIM2VR]{SIM2VR: Towards Automated Biomechanical Testing in VR}

\author{Florian Fischer}
\authornote{Both authors contributed equally to this research.}
\orcid{0000-0001-7530-6838}
\affiliation{%
 \institution{University of Cambridge}
 \country{United Kingdom}
 }

\author{Aleksi Ikkala}
\authornotemark[1]
\orcid{0000-0001-9118-1860}
\affiliation{%
  \institution{Aalto University}
  \country{Finland}
  }

\author{Markus Klar}
\orcid{0000-0003-2445-152X}
\affiliation{%
 \institution{University of Glasgow}
 \country{Scotland}
 }

\author{Arthur Fleig}
\orcid{0000-0003-4987-7308}
\affiliation{%
 \institution{ScaDS.AI Dresden/Leipzig} %
 \country{Germany}
 }
 
\author{Miroslav Bachinski}
\orcid{0000-0002-2245-3700}
\affiliation{%
 \institution{University of Bergen}
 \country{Norway}
 }
 
\author{Roderick Murray-Smith}
\orcid{0000-0003-4228-7962}
\affiliation{%
  \institution{University of Glasgow}
  \country{Scotland}
  }

\author{Perttu H\"{a}m\"{a}l\"{a}inen}
\orcid{0000-0001-7764-3459}
\affiliation{%
  \institution{Aalto University}
  \country{Finland}
  }

\author{Antti Oulasvirta}
\orcid{0000-0002-2498-7837}
\affiliation{%
  \institution{Aalto University}
  \country{Finland}
  }

\author{J\"{o}rg M\"{u}ller}
\orcid{0000-0002-4971-9126}
\affiliation{%
 \institution{University of Bayreuth}
 \country{Germany}
 }

\renewcommand{\shortauthors}{Fischer and Ikkala, et al.}

\begin{abstract}

Automated biomechanical testing has great potential for the development of VR applications, as initial insights into user behaviour can be gained {\it in silico} early in the design process. 
In particular, it allows prediction of user movements and ergonomic variables, such as fatigue, prior to conducting user studies. 
However, there is a fundamental disconnect between simulators hosting state-of-the-art biomechanical user models and simulators used to develop and run VR applications. 
Existing user simulators often struggle to capture the intricacies of real-world VR applications, reducing ecological validity of user predictions.
In this paper, we introduce \textsc{sim2vr}, a system that aligns user simulation with a given VR application by establishing a continuous closed loop between the two processes.
This, for the first time, enables training simulated users directly in the same VR application that real users interact with. %
We demonstrate that \textsc{sim2vr} can predict differences in user performance, ergonomics and strategies in a fast-paced, dynamic arcade game. In order to expand the scope of automated biomechanical testing beyond simple visuomotor tasks, advances in cognitive models and reward function design will be needed.

\end{abstract}



\begin{teaserfigure}
  \centering
  \includegraphics[width=\textwidth]{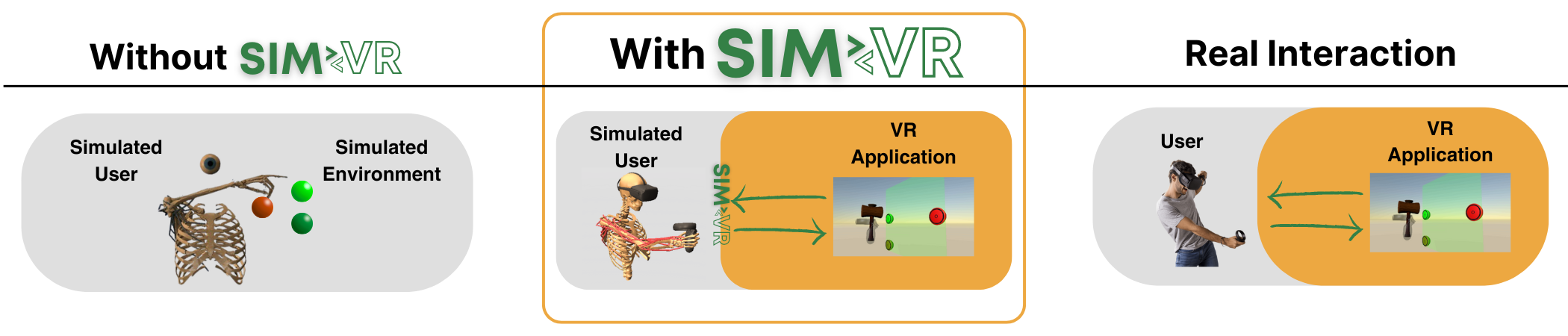}
  \caption{
  Simulating interactive user behaviour in VR typically requires re-implementing the game dynamics, logic and graphics of the VR application in another simulation engine, which is time-consuming, redundant and error-prone (left). \\
  \textsc{sim2vr} (middle) enables VR designers, for the first time, to run state-of-the-art biomechanical user simulations directly in the same VR application real users are interacting with (right).
  By aligning simulated and real interaction with respect to VR input and output, as well as application dynamics, \textsc{sim2vr} allows for more ecologically valid predictions of user performance, effort and strategies, supporting developers in the early stages of the design process.
  }
  \Description{Three boxes illustrating the simulated interaction loops without SIM2VR and with SIM2VR together with the real interaction loop. The "Without SIM2VR" loop (left) shows a biomechanical model with one eye ("Simulated User") and three targets in the same environment ("Simulated Environment"). The "With SIM2VR" loop (middle) shows a biomechanical model equipped with VR controller and HMD ("Simulated User"), a "VR Application" block, and the SIM2VR logo in between. An orange frame highlights this loop. The "Real Interaction" loop (right) shows a real user ("User") and the same "VR Application" block.}
  \label{fig:sim2vr-alignment}
\end{teaserfigure}

\maketitle

\section{Introduction}

Recent advances in deep reinforcement learning (deep RL) and biomechanical models have enabled the creation of \textit{simulated users} for modelling how humans may behave in interaction tasks\added{~\cite{Nakada18, Fischer21, Ikkala22, caggiano2022myosuite}}.
Using advanced biomechanical models that capture the intricacies of human body movement beyond linked-segment models, such simulated users allow prediction of time-continuous movements and effort-related quantities such as muscle energy expenditure~\cite{Umberger03} and fatigue~\cite{Xia08, Rashedi15}. %
In HCI, simulated users have been recently demonstrated for various tasks, including pointing \cite{Klar22, Cheema20}, typing \cite{Hetzel21, Jokinen21}, tracking \cite{Fischer21}, steering and choice reaction \cite{Ikkala22}. This modelling approach is promising, as it enables designers and developers to pose ``what-if?'' questions to assess how changes in design choices affect users' performance and ergonomics.

This bears particular promise for HCI domains where human data is scarce or costly to collect, such as in virtual reality (VR) interactions. In VR development, testing prototypes typically involves wearing a headset, standing up and moving around in space, which incurs significant overhead and renders iterative testing slow and cumbersome \cite{liu2023challenges}. Biomechanical user simulations show significant potential to remedy this as a means of \textit{automated testing}. 
In particular, they allow VR designers to evaluate and compare prototypes without the need for expensive user studies, which can be lengthy and stressful, especially for vulnerable participants.
By using an appropriate range of different user models, a simulation-based approach can increase diversity during the testing process, thereby promoting ability-based design~\cite{murraysmith2022simulation}.

To realise the potential of simulated users, we need to address the challenge of integrating model-based simulations into VR development workflows. State-of-the-art biomechanical models are not readily available in commonly used VR development environments like Unity,\footnote{\url{https://unity.com/}} or they cannot be combined with deep RL approaches to generate forward simulations in those environments. For instance, while there exists a MuJoCo Unity plugin\footnote{\url{https://github.com/google-deepmind/mujoco/tree/main/unity}} that provides access to biomechanical models implemented in MuJoCo~\cite{Todorov12}, an efficient physics engine suitable for biomechanical simulation, the models can only be controlled manually using the Unity Editor.

As a result, VR designers who wish to use automated testing must recreate the dynamics and graphics of the VR application within the simulation engine used for the biomechanical model. This, however, is impractical for three reasons. First, it introduces implementation overhead, in that two VR environments need to be maintained and updated in parallel. Second, building the replica is often tedious and cumbersome, and requires detailed knowledge of the subtleties of the considered application.
Third, the resulting simulation is unlikely to create ecologically valid predictions of interactive user behaviour, as the replica will typically correspond to a low-fidelity version of the original VR application. 
Such an approach thus considerably complicates \deleted{alignment of the user input sensed by the application, the (visual) output generated by the application, and the underlying application dynamics themselves}\added{the reduction of differences between the simulated interaction loop and the real interaction loop, which can be described by means of several \textit{alignment problems}, as detailed in Section~\ref{sec:idea-requirements}.}
\deleted{We call this the \textit{VR Simulation Alignment Problem} and describe it in detail in Section~\ref{sec:idea-requirements}.}
Simulations that do not adequately address \deleted{this problem}\added{these problems}
will provide inferior predictions of how real users interact with VR interfaces or techniques, limiting the validity of in silico testing and compromising the comparative analysis of design choices.

To overcome these limitations, we propose \textsc{sim2vr}, a system that integrates %
biomechanical user simulations directly into a given Unity application.
Alignment is improved by simulating hardware within the physics engine used for biomechanical simulation, tracking virtual VR input and output using OpenXR, accessing all relevant application data directly in Unity, and synchronising the user simulation and the VR application.
From a research perspective, this enables a more accurate and reliable development and validation of user simulation models by ensuring that \textit{real and simulated users interact with exactly the same virtual environment}\added{ (see Figure~\ref{fig:sim2vr-alignment})}. %
From an interactive software development perspective, our work provides a new solution to the long-standing problem of automated testing of VR interaction. %

To summarise, this paper contributes
\begin{itemize}
    \itemdeleted{ a novel definition and comprehensive discussion of the VR Simulation Alignment Problem encountered when simulating movement-based interaction in VR (Section~\ref{sec:idea-requirements});}
	\item \textsc{sim2vr}, a system \deleted{for integrating biomechanical simulations into Unity applications, which partially addresses the VR Simulation Alignment Problem}\added{that for the first time integrates biomechanical simulations directly into VR application development} (Section~\ref{sec:sim2vr-implementation}); %
	\itemdeleted{ a demonstration of SIM2VR's capability to predict differences in performance and ergonomics, and to reveal potential user strategies, verified by an accompanying user study (Section~\ref{sec:whac-a-mole}).}
    \item \added{an empirical assessment of \textsc{sim2vr}'s predictive capabilities for automated testing (Section~\ref{sec:whac-a-mole});}
    \item \added{a critical discussion on the scope as well as current and inherent limitations of \textsc{sim2vr} (Section~\ref{sec:discussion-limitations}).}
\end{itemize}
Open-source code of \textsc{sim2vr} is available at \url{https://github.com/fl0fischer/sim2vr}.

\section{Background and Context}\label{sec:related-work}
Below, we situate our work in relation to previous efforts in both automated testing in VR development, and computational user modelling and simulation.%

\subsection{VR Development and Automated Testing}

Because of the effort and time needed to don and doff equipment during prototype development, iterative testing and development in VR has a higher cost than traditional interface design for desktop or mobile software, making it typically slow and cumbersome \cite{liu2023challenges}. This underlines the need to automate testing as much as possible. \textit{Automated testing} allows prototypes to be automatically evaluated and compared without the need for expensive user studies~\cite{bierbaum2003automated, karakaya2022automated}, by predicting performance estimates such as completion times and error rates. 

While the need for automated testing has been clear for decades, a recent study of over 300 VR projects found that 79\% of the projects did not use any automatic tests, and that test automation followed common software testing practices without considering the user's body movements, focusing on aspects such as evaluating the correctness of a response after an event was triggered \cite{rzig2023virtual}. So far, even automated testing approaches designed specifically for VR do not emulate or simulate the user's moving body. Instead, they focus on moving and rotating the viewpoint to inspect VR scenes \cite{wang2022vrtest}, or operate on higher-level action events such as ``click UI button'' or ``grab object'' which abstract away the details of the user's movements \cite{bierbaum2003automated,harms2019automated, karakaya2022automated}. In the domain of Natural User Interfaces (NUIs), a generative motion model was proposed for automatic testing of gestural interaction in \cite{hunt2014automatic}, but unlike our work, the model was not embedded in a learning loop; instead, it generated random movement sequences without the ability to adapt to the tested interface.

\subsection{Computational User Modelling and Simulation}

\subsubsection{Simulation in HCI}
As discussed in \cite{murraysmith2022simulation}, user-centred design could move to a more rigorous, safer and predictable process via a simulation-based approach to interaction design. The same paper highlights general potential benefits of simulation in HCI for efficiency of design, safety, supporting diversity and transparency. Simulations can be used to predict task performance in interaction, such as time taken to finish the task and how often tasks can be successfully accomplished. For example, very early work on keystroke-level models in \cite{card1983psychology}, and more recent work \cite{kristensson2021design} used modelling to replace extensive experimentation, optimising text entry system parameters.

The general tools for simulation are improving rapidly, opening up new opportunities. 
As pointed out in \cite{lavin2021simulation}, we have seen recent developments in probabilistic, differentiable programming, high-performance computing and causal modelling. This, combined with the rapidly improving ability of machine learning to emulate complex aspects of human perception and behaviour (e.g. deep neural nets can simulate human-level visual perception \cite{he2016deep}, speech recognition \cite{graves2013speech} and control \cite{mnih2015human}), means that these new tools for simulation have an increased potential to be efficiently and usefully applied to domains such as HCI.

\subsubsection{Biomechanical User Simulation}\label{sec:rel-work_simulation-suites}
Biomechanical models and computer-based simulations are well-established %
\cite{Ayoub1974}. Increases in computational power have allowed them to evolve from simple models, limited to computation of mechanical loads in static postures~\cite{Winter1984}, to more physiologically-accurate musculoskeletal models~\cite{Delp07, Damsgaard06}. Biomechanical models were typically used for inverse simulation, e.g. using the OpenSim ecosystem~\cite{SETH2011212}. Such {\it inverse simulation methods}, namely inverse kinematics, inverse dynamics and static optimisation or computed muscle control, allow the estimation of mechanical loads within the human musculoskeletal system, and neural controls of the muscles given motion tracking data of a given user's movement as input ~\cite{Delp07}. %
In HCI, inverse dynamics have been used to identify fundamental ergonomics issues, such as ``gorilla arm'' \cite{jang2017modeling}.
{\it Forward simulation methods}, in contrast, are less frequently used in standalone situations, as they require muscle controls as inputs, which are extremely complex to measure experimentally. However, these have become more useful when used in conjunction with computational controllers~\cite{lee2016generating,Dembia2021,schumacher2023deprl} and efficient physics engines~\cite{Klar22}, which we will exploit in this paper.

\subsubsection{Optimal Feedback Control and Reinforcement Learning Models}
Control-theoretic models of human motion for mouse movements were developed in \cite{muller2017control} and supplemented by the assumption of \textit{Optimal Feedback Control}~\cite{Todorov02} in~\cite{Fischer22}. In such optimisation-based forward simulations, users are assumed to continuously interact with their environment by selecting controls such that a task-specific utility (e.g., a game score) is maximised, given the perceptual information available at decision time.
By including intermittent control behaviour typical of humans, simulations could also capture more realistic models of human motor variability~\cite{martin2021intermittent}. 

Reinforcement learning has recently emerged as a suitable framework for modelling human behaviour in a flexible way: one only needs to define the states, actions and rewards, and then use RL methods to estimate the optimal policy \cite{chen2015emergence, chen2017cognitive}. In some cases, the reward function and other parameters can be inferred from human data~\cite{Banovic16, Kangasraasio17}.
If the reward function and state-action space, including their key limitations, are similar to those of a human, increasingly human-like behaviour has been shown to emerge through learning \cite{oulasvirta2022computational}. 

Recent applications in HCI include models of typing, menu selection, multitasking and visual decision-making \cite{oulasvirta2022computational}. %
Using the assumptions of signal-dependent control noise and movement time minimisation, Fischer et al.~\cite{Fischer21} have shown that an RL agent can generate human-like mid-air trajectories that are in 
accordance with well-established phenomena such as Fitts' Law~\cite{Fitts54} and the Two-Thirds Power Law~\cite{Lacquaniti83}.
RL-based simulation may also provide valuable information for predicting usability- and ergonomics-related criteria and to aid in interface design. For instance, Cheema et al. trained a torque-actuated biomechanical arm model in a mid-air pointing task, and used the model to predict fatigue of real human subjects performing the task~\cite{Cheema20}.
In this paper, we build on these developments and use RL to model interaction in VR. %

\subsection{Creating Simulated Users}\label{sec:creating-simulated-users}
By combining perception models, musculoskeletal models, and physically simulated input devices, we can train RL agents, or simulated users, to model and simulate intricate interaction tasks, such as those requiring visuomotor control. An early example of such control was presented in~\cite{Nakada18}, where Nakada et al.\ introduced a virtual human model and used deep learning to learn reaching and tracking tasks.

The development of computationally efficient simulation software, such as MuJoCo~\cite{Todorov12}, has permitted the use of deep RL for flexible problem formalisations of interaction tasks. This allows a researcher to guide an agent's learning through reward functions, as was demonstrated in~\cite{Ikkala22} by Ikkala et al. with their \textit{User-in-the-Box} (UitB) approach, where they trained a muscle-actuated model of the upper torso and right arm to complete a series of interaction tasks. UitB allows combining biomechanical models with perception models and movement-based interaction tasks through a modular software structure, which facilitates integrating custom model components. %
In our work, we exploit this modularity and extend UitB to VR environments.

MyoSuite~\cite{caggiano2022myosuite}, another user simulator implemented in MuJoCo, focuses on dexterous hand, lower- and full-body movements. MyoSuite also provides a conversion tool for converting OpenSim models into MuJoCo format (MyoConverter\footnote{\url{https://github.com/MyoHub/myoconverter}}~\cite{wang2022myosim, Ikkala20}), which can also be readily added into the UitB framework.

Simulated users could promote ability-based design in VR, as diverse user populations, i.e., populations with different physical or cognitive properties, could be simulated without the risk of physical harm~\cite{murraysmith2022simulation,QueBolWil2011,quek2013phd}.
However, existing approaches that create and evaluate simulated users in the context of HCI are limited to ``mock-up'' interaction environments that require a re-implementation of the application dynamics real users face during interaction~\cite{Fischer21, Ikkala22, Hetzel21}.
These approaches are thus fundamentally unsuitable for automated biomechanical testing of a given VR prototype.
To the best of our knowledge, %
our work is the first to create simulated users in an ecologically valid VR context by training biomechanical user models directly in the same VR environment that human users interact with. %

\section{VR Simulation Alignment}
\label{sec:idea-requirements}

As depicted in Figure~\ref{fig:sim2vr-alignment} (right), movement-based interaction in VR can be regarded as a continuous-time closed loop between the two core components of interaction: the user and the VR application.
During this interaction, the user obtains multi-sensory information from \deleted{its}\added{their} environment\deleted{\footnote{Usually, this includes information from the peripersonal and physical space of the user (e.g., proprioceptive or haptic input signals) as well as the virtual space (visual input signals).}} and produces a neuromuscular control signal to generate body movements, hence manipulating the input device to complete the given interaction task.
The VR application, on the other hand, conveys an interactive, virtual environment that can be manipulated via one or multiple input devices. The state of the input device is actively sensed by the VR system, measuring, e.g., how the VR controller is moved, or whether a physical controller button is pressed. %
As a final step to close the loop, the VR application provides sensory output to the user, such as a rendered image displayed on an HMD, or haptic output in the form of temporary controller vibrations.

\deleted{The VR}\added{This} interaction loop \deleted{described above }not only characterises the movement-based interaction between a \textit{real} user and a given VR system, but can also be used as a framework for creating model-based \textit{simulations} of VR interaction.
To ensure that the simulated interaction loop acts as a reasonable proxy for real user interaction, i.e., to ensure external validity of the predictions made, \textbf{the two interaction loops need to be aligned}.
\deleted{We coin this problem the \textit{VR Simulation Alignment Problem}.} %

As \deleted{summarised in Table~\ref{tab:alignment-problems} and }described in detail below, \added{we categorise }VR simulation alignment \deleted{evolves in}\added{along} six dimensions, three of which focus on the alignment of the simulated user and three on the alignment of the simulated VR system. %
~ \\

\subsubsection*{Physical Alignment}
The physics of the interaction, i.e., the biomechanics of the user's body and the dynamics of the input and output devices, need to be captured adequately by the simulation. 
\deleted{This requires a sophisticated biomechanical model that captures the complexity of the human musculoskeletal system, at least for the body parts relevant for the considered interaction (e.g., a model of the shoulder, arm and hand), as well as simulated input and output devices (e.g., a simulated VR controller and an HMD) that align with the real hardware in terms of static and dynamic properties such as mass or size.}
\deleted{Too simple a model of the user's body (e.g., a linked-segment model of the arm) may produce unrealistic movement trajectories, resulting in severe deviations from how real users interact with a given application.} 

\subsubsection*{Input Channel Alignment}
The simulation needs to sense the state of the input device (e.g., the location and orientation of the VR controller) in the same way as the real VR system. %
In particular, it has to account for coordinate frame transformations %
and update rates (i.e., the measuring frequency).

\subsubsection*{Application Alignment}
The simulated VR application should have the same internal states as the target VR application (e.g., it needs to provide the same virtual objects and affordances) and show the same responsive behaviour, given that both applications are provided the same input signals.
\deleted{If application alignment is low, e.g., because a game constraint %
is missing or is incorrectly re-implemented, %
applying the same body movements in the simulation as in reality may result in substantially different states of the VR application.}

\subsubsection*{Output Channel Alignment}
The simulated VR system needs to provide the same output signals \added{(e.g., HMD image) }as the real VR system, given that both are in the same state. This ensures that the simulated user has the chance to perceive the same information from the application as a real user.
\deleted{In particular, the simulated user needs to be provided the same rendered image as is shown on the the real user's HMD.} %

\subsubsection*{Perceptual Alignment}
The simulated user should perceive \deleted{its}\added{their} environment (including \deleted{its}\added{their} own body, physical input devices and the VR application) in the same way \added{as }a real user \deleted{does}\added{would}. This requires appropriate modelling of all sensory modalities relevant for the considered interaction, including vision, proprioception and haptics. 
\deleted{Note that perceptual alignment focuses on the observations that users make of their environment, whereas output channel alignment is about the information that the VR application makes available to the user in the first place.}

\subsubsection*{Cognitive Alignment}
Given an interaction task (e.g., to successfully complete a specific game level as fast as possible), \deleted{the simulated user should generate the same muscle control signals as a real user, when provided with the same perceptions.}\added{the simulated user should have a similar internal representation and select the same control output when given the same sensory input.}
In the context of reinforcement learning, this requires defining an appropriate task-specific reward function such that an optimal control policy resembles real user behaviour.
~ \\

As the behaviour of the user and the VR application depend on each other in a closed loop manner, errors can easily cascade in time. This results in an increasing divergence between the simulation and the real VR interaction, once an error is introduced. For example, in the Whac-A-Mole game from Section~\ref{sec:whac-a-mole}, a different \deleted{grasp}\added{hammer grip} (e.g., due to a slight offset in the game dynamics) may result in fewer/more/different targets being hit, which has an impact on the movements generated by the simulated user at later times, leading to even greater differences in target hits, and so on. The same holds for a missing or incorrectly implemented game constraint: if the simulated user has to hit a target with less power or less often to receive the same game reward, \deleted{it}\added{they} will learn a totally different control policy, leading to even more divergent application states.
Depending on the complexity and the number of discontinuities in the interaction dynamics, these errors may pollute any variable of interest obtained from the simulation. %
It is thus essential to ensure alignment in as many dimensions as possible.

\section{SIM2VR: Enabling Biomechanical User Simulations in VR}\label{sec:sim2vr-implementation}

\begin{figure*}[h!]
	\centering
	\includegraphics[width=\textwidth]{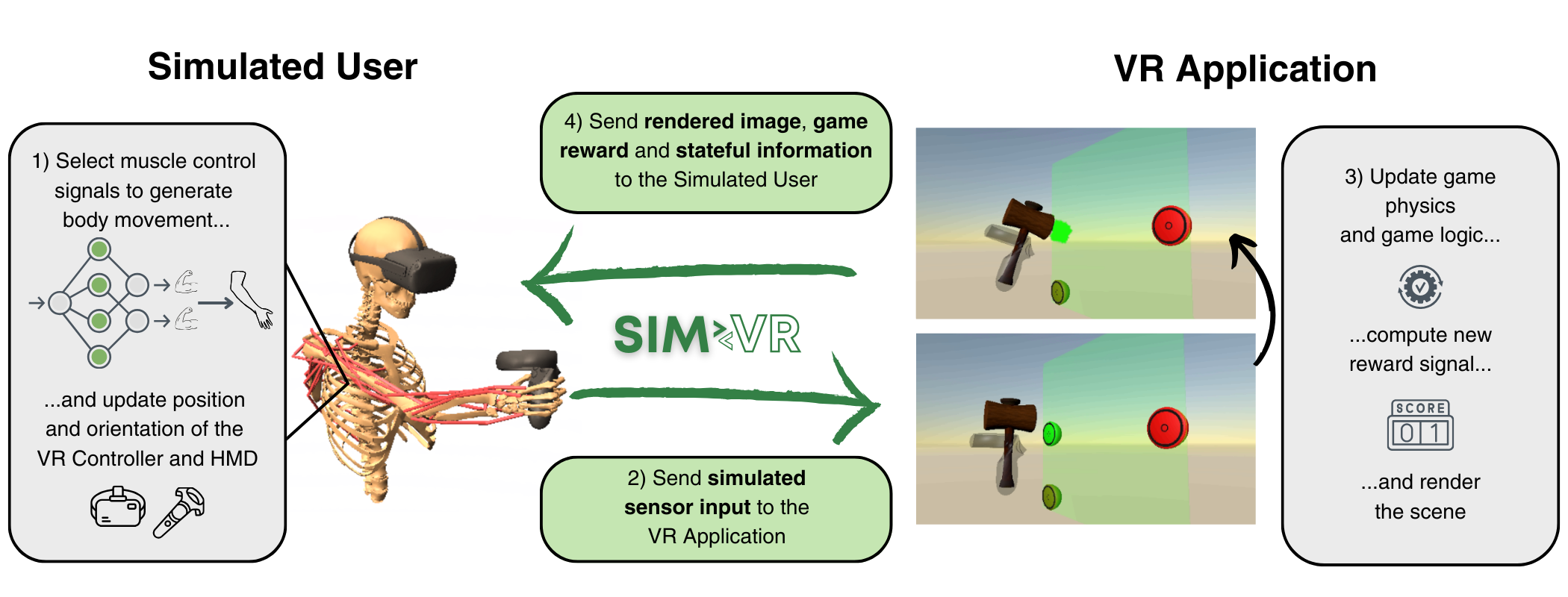}
	\caption{
		To obtain valid and reliable predictions of interactive user behaviour in VR, a simulated user needs to perceive and control exactly the same VR application as real users. Our \textsc{sim2vr} system addresses this \deleted{\textit{VR Simulation Alignment Problem} }by establishing a closed loop between the simulation process and the desired VR application, ensuring that the same input and output signals are generated as with real hardware. The resulting time-continuous integration allows to train and evaluate a biomechanical user simulation directly in a given VR environment.
	}
	\Description{A closed-loop between the simulated user (shown as biomechanical model with VR controller and HMD) and the VR application (visualised by two example frames from the Whac-A-Mole game) is established through the SIM2VR system (visualised by an arrow from the simulated user to the VR application and an arrow in the opposite direction, with SIM2VR logo in between). This loop consists of four steps (the following texts are annotated in corresponding text boxes): 1) [Simulated User] "Select muscle control signals to generate body movement and update position and orientation of VR controller and HMD." 2) [SIM2VR] "Send simulated sensor input to the VR application." 3) [VR Application] "Update game physics and game logic, compute new reward signal and render the scene." 4) [SIM2VR] "Send rendered image, game reward and stateful information to the simulated user."}
	\label{fig:simulated-interaction-loop}
\end{figure*}

In this section, we propose \textsc{sim2vr}, a system that allows two simulators, the simulated user and the original VR application, to run in parallel while being synchronised temporally and spatially.
The aim of this system is to establish a closed interaction loop between the simulated user and the VR application (see Figure~\ref{fig:simulated-interaction-loop}) that closely resembles the interaction loop between the real user and the VR application\added{. While aligning the behaviour of the simulated user with that of real users in all dimensions constitutes a highly complex task, \textsc{sim2vr} addresses the system-related alignment problems (input channel, application, and output channel alignment)} described in Section~\ref{sec:idea-requirements}. 

\deleted{In this simulated interaction loop, t}\added{The core idea is as follows: }the simulated user generates muscle control signals based on \deleted{its}\added{their} sensory perception of the body, the input devices, and the VR application (e.g., visual information of the scene and proprioceptive information of the body posture).
These control signals are used to actuate the muscles, driving the movement of the body, and, consequently, the simulated states of the hardware devices.
The position and rotation of the VR controllers and HMD are then measured by virtual sensors and sent to the VR application. 
The VR application processes this input, computes a reward based on the updated application state (e.g., the current game score) and renders the image perceived through the virtual HMD.
Finally, the rendered image is provided to the simulated user, along with the reward and optional ``stateful'' information (such as time remaining in an interaction task), thus closing the interaction loop.\deleted{\footnote{Note that the reward signal and the stateful information are only included in the \textit{simulated} interaction loop. This is mainly for technical reasons: The reward signal is required for an RL method to learn a control policy, and the stateful information can be used to improve the training.}}

\deleted{Crucially, this approach ensures application alignment by giving the simulated user access to exactly the same VR application that real users interact with. Based on this approach, we developed the \textsc{sim2vr} system, which additionally addresses the input channel and output channel alignment problems to provide more ecologically valid predictions of VR interaction.}

\added{Crucially, this approach ensures application alignment by giving the simulated user access to exactly the same VR application that real users interact with. This is contrary to existing biomechanical frameworks, such as User-in-the-Box (UitB)~\cite{Ikkala22} or MyoSuite~\cite{caggiano2022myosuite}, which are not designed for modelling interaction with real-world VR applications. These frameworks typically run a physics engine that is incapable of capturing the internal states, dynamics, shapes and visuals of a VR application, which fundamentally limits their ability to reliably predict user behaviour.
We intentionally refrained from re-implementing the target VR environment in one of these simulation framework for the purpose of this paper, as this would impose considerable implementation and maintenance overheads, and ultimately result in a low-fidelity replica that is inevitably inferior to the VR environment real users are interacting with. %

}

\subsection{Implementation Principles \& Details}\label{sec:implementation}

\added{We have implemented }\textsc{sim2vr} \deleted{is}\added{as} an extension of UitB to Unity\footnote{\added{\url{https://unity.com/}}} applications, \deleted{which we built }based on the following principles \added{(}outlined in the remainder of this section\added{)}:
\begin{itemize}
    \item Simulated Hardware
    \item Standardised VR Interaction Library
    \item Direct Access to the VR Environment
    \item Synchronised Simulators
\end{itemize}

We opted to use Unity, as it provides an extensive set of VR/AR features and has a large community of VR developers. 
The VR application in the simulated interaction loop can be any existing Unity application or a prototype thereof, either built as a standalone app or run within the Unity Editor.

\begin{figure}[h!]
    \centering
    \includegraphics[height=0.2\textheight]{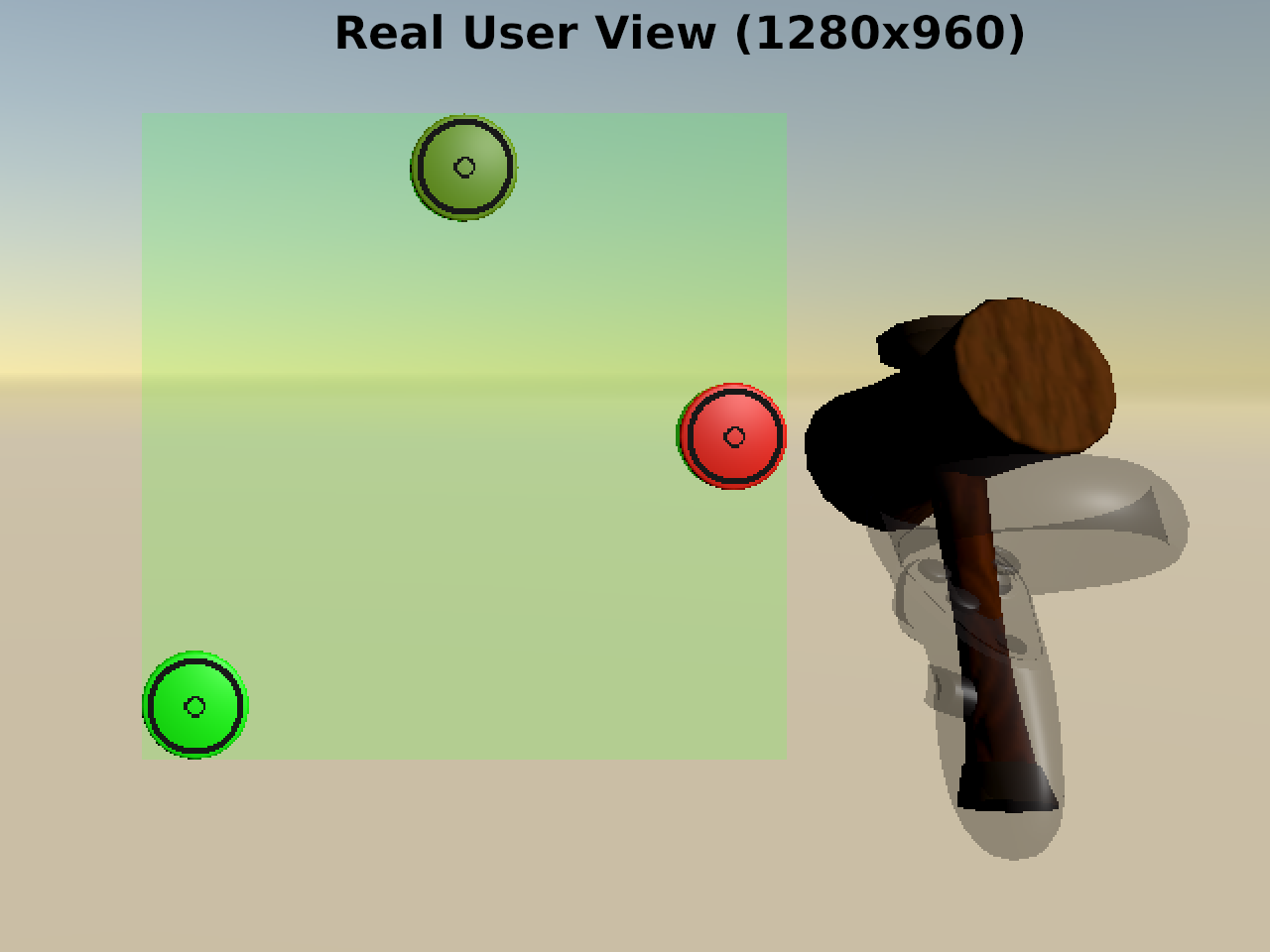}
    \includegraphics[height=0.2\textheight]{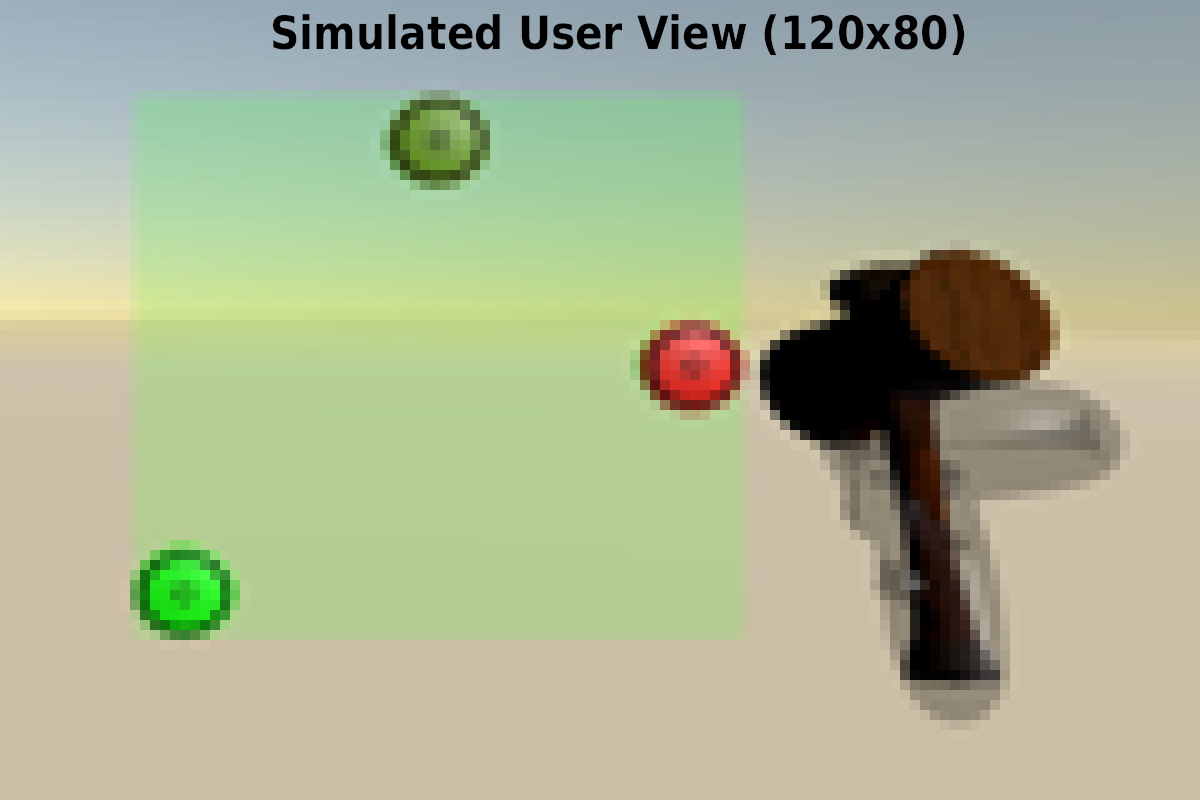}
    \caption{With \textsc{sim2vr}, simulated users \deleted{for the first time }have access to exactly the same visual output as humans. The top figure depicts how real users perceive a VR environment %
    (in this case, the Whac-A-Mole game from Section~\ref{sec:whac-a-mole}), while the bottom figure shows the downsampled RGB-D image that the simulated user perceives.}
    \label{fig:visual-output-diff}
    \Description{Sample renders of the Whac-A-Mole game as shown to real users (top; annotation: "Real User View (1280x960)") and the simulated user (bottom; annotation: "Simulated User View (120x80)"), respectively.}
\end{figure}

The simulated user can be created within UitB, which requires selecting a biomechanical model implemented in MuJoCo, as well as one or more perception modules. 
These perception modules define how the simulated user perceives \deleted{its}\added{their} surroundings, e.g., via visual or proprioceptive signals.
To model how users perceive the virtual environment when wearing an HMD, \textsc{sim2vr} adds a novel vision module denoted as \textbf{\texttt{UnityHeadset}} to UitB.
\deleted{This module enables the simulated user to observe the rendered image sent to the virtual HMD as an RGB-D array (see Figure~\ref{fig:visual-output-diff}).} 
\added{This module enables the simulated user to observe the Unity scene through the same \texttt{Camera} game object that renders the scene to real users (see Figure~\ref{fig:visual-output-diff}). The rendered images are presented to the simulated user as 120$\times$80 pixel RGB-D arrays. The module also allows the visual observation to be filtered and enhanced, for example by omitting certain colour channels or stacking prior observations.}

\begin{figure*}[h!]
    \centering
    \includegraphics[width=\textwidth]{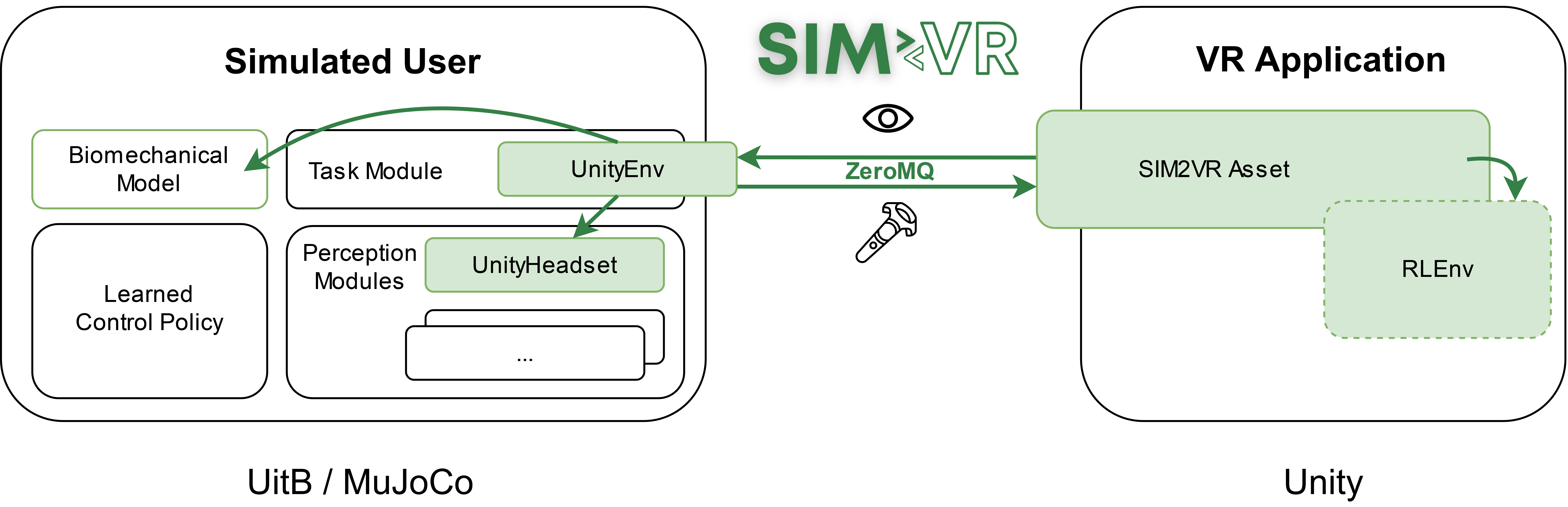}
    
    \caption{\textsc{sim2vr} enables efficient data transfer between the simulated user (implemented in UitB) and the VR application (implemented in Unity). This is achieved by establishing a ZeroMQ connection between the generic \texttt{UnityEnv} task module and the \texttt{SIM2VR Asset}, both provided by \textsc{sim2vr}. The \texttt{UnityEnv} equips the biomechanical model with VR controllers and an HMD. Visual perception of the VR application is received through the \texttt{UnityHeadset} vision module, based on the rendered HMD images sent to the \texttt{UnityEnv}. The \texttt{SIM2VR Asset} also allows application-dependent rewards, such as game scores, to be defined directly in the VR application using the \texttt{RLEnv} Unity script. The rewards can be used by the simulated user to improve \deleted{its}\added{their} control policy while interacting with the VR application.
    }
    \Description{Components of the SIM2VR platform. The "Simulated User" consists of a "Biomechanical Model", a "Task Module", a "Learned Control Policy" and one or multiple "Perception Modules", and is implemented in the UitB framework (which makes use of MuJoCo). The "VR Application" is implemented in Unity. SIM2VR connects these two components through a ZeroMQ connection that allows to transfer sensor input data (symbolised by a VR controller icon) from the "Simulated User" to the "VR Application", and visual perceptions (symbolised by an eye icon) in the opposite direction. This is achieved by adding the "UnityEnv" task class and the "UnityHeadset" vision module to the simulated user, and the "SIM2VR Asset" to the VR application. The latter also provides a template for setting up the application's "RLEnv". Additional arrows are drawn from "UnityEnv" to "Biomechanical Model", from "UnityEnv" to "UnityHeadset" and from "SIM2VR Asset" to "RLEnv". The "RLEnv" block has a dashed border.}
    \label{fig:sim2vr-components}
\end{figure*}

In the UitB framework, the interaction environment and the rewards that incentivise completing the given interaction task need to be defined in a so-called task module.
To enable using Unity applications as interaction environments, \textsc{sim2vr} provides a \textbf{\texttt{UnityEnv}} task module on the UitB side, and a \textbf{\texttt{SIM2VR Asset}} on the Unity side. These are connected via ZeroMQ, which creates a communication channel between the two simulators. \added{As \texttt{UnityEnv} handles the communication, it first receives the rendered images from the virtual HMD, before relaying them to the \texttt{UnityHeadset}. Later, after the simulated user's pose has been updated, \texttt{UnityEnv} sends the HMD and controller \deleted{information}\added{position and rotation} to the \texttt{SIM2VR Asset}.}
A detailed overview of the components of \textsc{sim2vr} and their dependencies is given in Figure~\ref{fig:sim2vr-components}.

The following principles are essential for the functionality of \textsc{sim2vr}:

\subsubsection*{Simulated Hardware}
The simulated user needs to be equipped with VR controllers and an HMD in order to be able to interact with the VR application in the same way as a real user.
To this end, the \texttt{UnityEnv} task module augments the biomechanical MuJoCo model of the simulated user with respective hardware objects.\footnote{The module currently comes with mesh files of the Meta Quest 1 \& 2 (\href{https://www.meta.com/quest/}{https://www.meta.com/quest/}), however, these can be easily replaced with models of other VR equipment.} 
In particular, it ensures that the VR controllers and HMD are rigidly attached to the hands and head of the biomechanical model. 

\subsubsection*{Standardised VR Interaction Library}
To achieve input channel alignment, the VR application needs to sense the location and orientation of simulated hardware just as if a physical controller and HMD were used.
\textsc{sim2vr} enables such hardware-agnostic input handling by building on the OpenXR\footnote{\url{https://www.khronos.org/openxr/}. \textsc{sim2vr} requires the Unity application to handle interaction with VR devices using the OpenXR plugin (version 1.5.3 or later).} standard.

\subsubsection*{Direct Access to the VR Environment}
By adding the \texttt{SIM2VR Asset} to the desired Unity application, \textsc{sim2vr} gains direct and genuine access to all Unity variables relevant for the interaction.
This has several advantages. First, it permits capturing the output signals, such as the rendered image displayed on the HMD, and directly forwards these signals to the simulated user.\footnote{While \textsc{sim2vr} is currently limited to the transmission of visual output signals, we plan to support other feedback modalities such as auditory and haptic output.}
It is thus key in ensuring output channel alignment.
Second, it can be used to define application-specific rewards, e.g., based on a game score. This is done in the \textbf{\texttt{RLEnv}} class provided by the \texttt{SIM2VR Asset}, \added{which must be inherited and defined separately for each Unity application to ensure compatibility with RL training.}
Third, it allows the improvement of  the RL training by providing further stateful information (e.g., the time passed in an episode, or the number of completed subtasks) as observations to the simulated user, and to log\deleted{\footnote{To facilitate troubleshooting, we have added \textit{Weights \& Biases} (\url{https://wandb.ai/site}) integration, which allows real-time monitoring of the simulated user's performance.}} any Unity variables of interest (e.g., the current performance in terms of specific subtasks).

\subsubsection*{Synchronised Simulators}
The physics engine of the user simulation and the VR application need to be temporally synchronised.
\textsc{sim2vr} achieves this by including the current as well as the next timestamp of the user simulation as metadata in the data packages sent to Unity. These timestamps are used to halt the VR application updates until the same amount of time has elapsed as in the user simulation. 
To avoid computational overhead, the Unity scene is rendered exactly once per UitB time step, and only relevant data is transferred between the two simulators. Faster than real time simulations are achieved by modifying the Unity \texttt{timescale} parameter (in our experiments we ran Unity at 5$\times$ real time).
The \texttt{UnityEnv} module can also trigger the reset function of the \texttt{SIM2VR Asset} and vice versa. This is essential to ensure that both the simulated user and the VR application are reset together when necessary, for example, to start a new episode during RL training, or after completing a game level. \added{The simulators are also spatially synchronised by means of aligning the simulators' coordinate systems and appropriately transforming coordinates between them.}

\added{
\subsection{Creating User Simulations in Unity: A Step-by-Step Guide}\label{sec:step-by-step}
In the following, we first describe how \textsc{sim2vr} can be used to generate user simulations for a given Unity application. Then, following the guide, we demonstrate how \textsc{sim2vr} was used to train a simulated user to play a ``Beat Saber'' style game implemented in the \textit{VR Beats Kit}, a third party software freely available on the Unity Asset Store.\footnote{\href{https://assetstore.unity.com/packages/templates/systems/vr-beats-kit-168243}{https://assetstore.unity.com/packages/templates/systems/vr-beats-kit-168243}} A more detailed version of this walkthrough is available at the project website \url{https://github.com/fl0fischer/sim2vr}.

\subsubsection{Importing the Asset}
First, the \texttt{SIM2VR Asset} must be imported into the Unity project. After adding the \textit{sim2vr} prefab as a game object into the desired scene, the \textit{SimulatedUser} game object needs to be connected to the \textit{VR Controllers} and \textit{Main Camera} provided by the OpenXR plugin. The \texttt{SIM2VR Asset} contains \textit{Recorder} and \textit{Logger} game objects, which can be used to record game play and log controller and headset poses, as well as further game data, during training and evaluation. Due to the reliance on game-specific data, the \textit{Logger} must be adapted to each game separately.

\subsubsection{Defining the Reward Function}

The majority of reward functions that have been used successfully to date have some common features. Typically, they consist of a task-specific component that incentivises reaching the considered goal(s) (e.g., reaching a target position, maintaining a body posture, or moving at a certain speed) and effort costs.
Effort costs usually act as a regularisation, i.e., they constrain the set of possible solutions by favouring policies that achieve the goal with less effort.
The task-specific reward can be either sparse (i.e., reward is only given after completing the task) or dense (i.e., rewards are continuously given, depending on how ``close'' the agent is to completing the task). Dense rewards are preferable from a mathematical point of view, as they provide smooth gradients, guiding the RL agent towards the optimal policy. However, defining the ``right'' dense rewards for a given interaction task, i.e., rewards that incentivise task completion but are not biased towards a particular strategy, can be quite challenging. We provide a \textit{reward scaling tool}, which can be used as a starting point for defining such rewards; the tool is described in the supplementary material.

In \textsc{sim2vr}, the task-specific reward needs to be calculated and stored within the \textit{CalculateReward} method of \texttt{RLEnv}. If the VR application under consideration provides a game score, this score can be used directly to define the task-specific reward.\deleted{\footnote{Note that game scores typically accumulate points throughout the episode, so the reward signal should correspond to the difference in score since the last frame.}} The effort costs can be defined when configuring the UitB user model.

\subsubsection{Defining the Reset Functionality}

The \texttt{RLEnv} class contains methods that need to be implemented to reset the game to an initial state at the end of each episode. This usually includes destroying game objects created during runtime and resetting variables used to compute the reward. All code related to resetting the reward variables should be defined in the \textit{InitialiseReward} method. Preparations for the next episode, such as choosing a game level, can be defined in the \textit{Reset} method, or, if these should be called only once when starting the game, in the \textit{InitialiseGame} method. The end of an episode is indicated by the \textit{UpdateIsFinished} method.

\subsubsection{Defining the Stateful Information (optional)}
Since including an application- and task-dependent time feature as stateful information in the observation can help to train the RL agent, the \texttt{RLEnv} class provides a \textit{GetTimeFeature} method that allows time information to be appended to the observation of the simulated user. Note that in deterministic games, the inclusion of such a time feature may lead to a control policy that exploits this information instead of relying on (visual) observations.

\subsubsection{Defining the Simulated User}

After preparing the Unity application for running user simulations, an appropriate user model needs to be created in UitB. This mainly involves the choice of a biomechanical user model (including effort cost) as well as perception modules beyond the default \texttt{UnityHeadset} vision module, both of which can be easily selected within the UitB config file.
Other optional parameters in the config file include the position and orientation of the HMD relative to a specific body part of the biomechanical user model, arguments to be passed to the Unity application (e.g., setting game level/difficulty), and RL hyperparameters (e.g., network size, time steps, batch size, etc.).

To ensure that all the target positions required by the VR application can be reached by the biomechanical model, we provide a \textit{reach envelope tool}. This tool allows interactive visualisation of the body-centred positions that a given simulated user can theoretically reach, along with the target positions that occur in a given VR interaction task; the tool is described in the supplementary material.

\begin{figure*}[!h]
    \centering
    \raisebox{1.75cm}{\subfloat[Game View]{\includegraphics[width=0.19\textwidth]{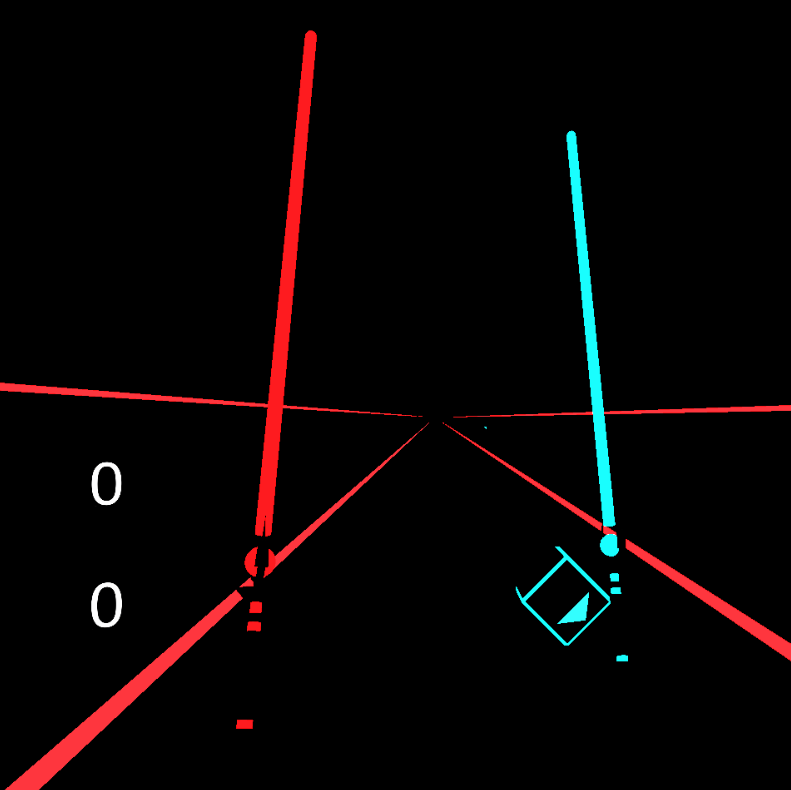}\label{fig:beatsvr-game-view}} \quad}
    \unskip\ \vrule\
    \raisebox{1.5cm}{\subfloat[Simulated User]{\quad \includegraphics[width=0.72\textwidth]{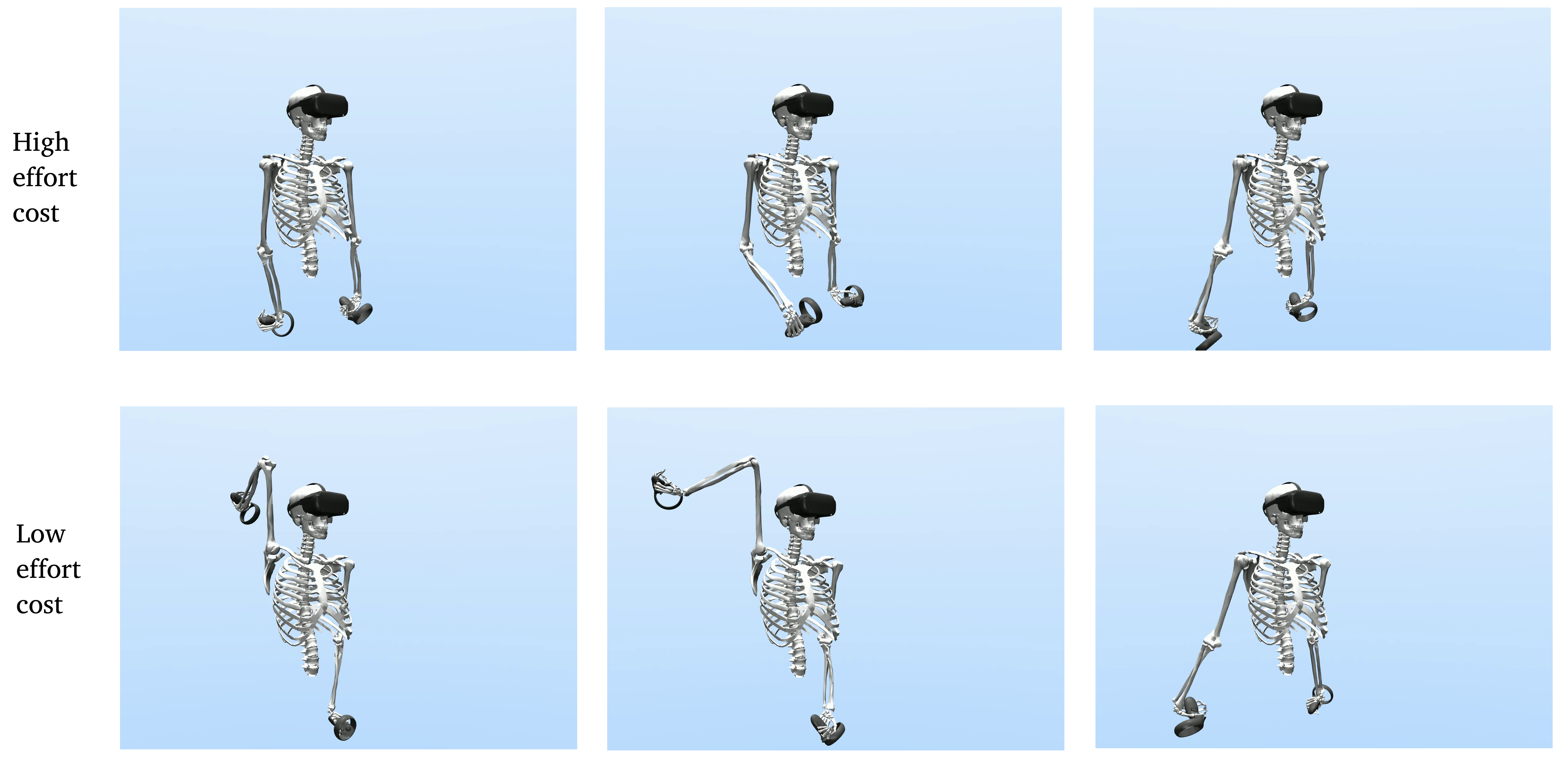}\label{fig:beatsvr-example}}}
    \caption{
    a) In the VR Beats game, the player controls two lightsabers and has to cut through boxes that are moving towards them. %
    b) Still frames from evaluations of trained simulated users playing the VR Beats game, about to hit an incoming box. Top: A simulated user trained with a high effort cost weight ($0.05$) has learned to keep \deleted{its}\added{their} arms down and rely on wrist movements. Bottom: A simulated user trained with a low effort cost weight ($0.001$) moves \deleted{its}\added{their} arms considerably more, sometimes resulting in implausible postures.
    }
    \Description{a) A scene from the VR Beats game shows a red and a blue lightsaber. A blue box is moving towards the view point on the lower right side. Two zeros on the left show the game score. b) A total of 6 figures displayed in two rows and three columns. The top row (label: "High effort cost") shows a skeletal human model performing a hitting motion while keeping the arm down and relying on wrist movement. The bottom row (label: "Low effort cost") shows the same model performing a hitting motion with a raised arm.}
\end{figure*}

\subsubsection{Example: VR Beats Kit}
In VR Beats, players wield lightsabers in both hands and attempts to hit all incoming boxes while dodging walls and avoiding mines\added{ (see Figure~\ref{fig:beatsvr-game-view})}. Since the biomechanical user models available in UitB cannot move their torso, we trained the simulated user to play only the first 10 seconds of the game, which does not include any walls (or mines). During the first 10 seconds, a total of 4 boxes pass the player, and missing a box causes the episode to end. A successful hit is rewarded with a score between 50-200 points, depending on the speed and direction of the hit. Below we describe the definition of each functionality required by \textsc{sim2vr}.

\paragraph{Reward Function}
As a task-specific reward, we used the original (sparse) game score (calculated based on the number of correct and incorrect target hits), as it was easy to implement and not prone to unwanted biases. We also used the default \textit{neural} effort cost implemented in UitB.

\paragraph{Reset Functionality}
For this game, it is sufficient to invoke the existing \textit{onGameOverEvent} to signal that the episode has ended, destroy the existing game objects, and then invoke the \textit{onRestart} event to restart the game. The reward is initialised in the \textit{Reset} method. To check if the episode has terminated, we make use of the existing \textit{getIsGameRunning} method of the \textit{VR\_BeatManager} instance.

\paragraph{Stateful Information}
No additional stateful information was provided to the simulated user.

\paragraph{Simulated User}
As VR Beats spawns targets for both hands, we implemented a bimanual UitB user model by converting a bimanual version of the \textit{MoBL ARMS model}\footnote{\url{https://simtk.org/frs/?group_id=657}}~\cite{Saul14} using the MyoConverter~\cite{Ikkala20,wang2022myosim}. To speed up the training process, we replaced the muscle actuators with torque actuators attached directly to the joints.
As perception modules, we used the default UitB proprioception module and the \texttt{UnityHeadset} vision module provided by \textsc{sim2vr}. The former allows the simulated user to infer joint angles, velocities and accelerations, as well as muscle activations and index finger position. The latter is configured to use the red colour and depth channels of the RGB-D image, and to stack the current observation with a delayed (0.2 seconds prior) visual observation; this configuration allows the control policy to distinguish between left and right arm targets, and to infer target movement.

\paragraph{Results}
We trained three simulated users with neural effort cost weights $0.001$, $0.01$ and $0.05$, respectively. These were chosen based on the suggestions provided by the \textit{reward scaling tool}. These cost weights allow us to explore how the behaviour of the simulated user can vary from an ``explorative'' or ``enthusiastic'' player to a more ``lazy'' player.
For instance, the simulated user trained with the highest effort cost weight ($0.05$) learned to do only the bare minimum to move \deleted{its}\added{their} arms and hit the incoming targets. In particular, \deleted{it}\added{they} hit the targets by relying mostly on wrist movements, suggesting that the game design could benefit from modifying the target box trajectories to encourage larger arm movements. The simulated users trained with lower effort costs learned to move their arms more, with the lowest effort leading to a policy where the arm movements are exaggerated. Figure~\ref{fig:beatsvr-example} shows still frames of simulated users trained with either high or low effort cost.}

\section{Predicting Interactive User Behaviour with SIM2VR}\label{sec:whac-a-mole} 
In this section, we show how \textsc{sim2vr} can be used to predict movement-based interaction in VR and demonstrate its utility in evaluating VR interface design choices.
We \added{therefore} developed a VR version of the popular, visuomotorically demanding, fast-paced and dynamic arcade game \textit{Whac-A-Mole} in Unity with different interface variations. 
\added{This allows us to investigate in a controlled way the extent to which \textsc{sim2vr} can predict the effect of design choices such as difficulty or target area placement.}
With \textsc{sim2vr}, we train a UitB model to play the game, and then use this model to predict differences in performance, effort and strategies between different game variations. To assess the quality of these predictions, we compare them to reference data from a user study, where real humans were asked to play the same VR game.

\deleted{We re-implemented our own version of the Whac-A-Mole game, as this allows us to investigate in a controlled way the extent to which \textsc{sim2vr} can predict the effect of design choices such as difficulty or target area placement. In the supplementary material, we show how to integrate \textsc{sim2vr} into an existing (third-party) Unity application and discuss how to create a simulated user in UitB.}%

\begin{figure}[h!]
	\centering
	\subfloat[Front View with Target Grid]{\includegraphics[width=0.45\textwidth]{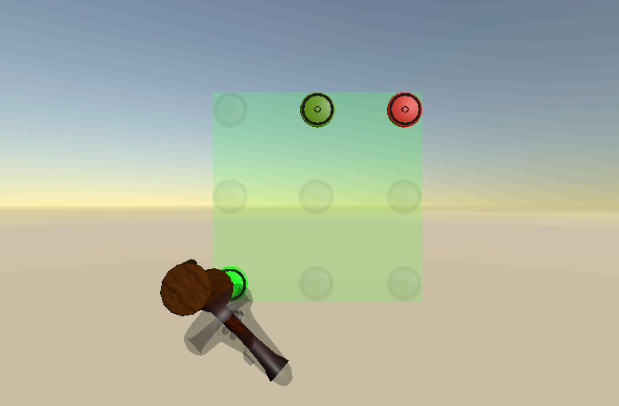}\label{fig:whac-a-mole-front}}
	\hfill
	\subfloat[Target Area Placements]{\includegraphics[width=0.45\textwidth]{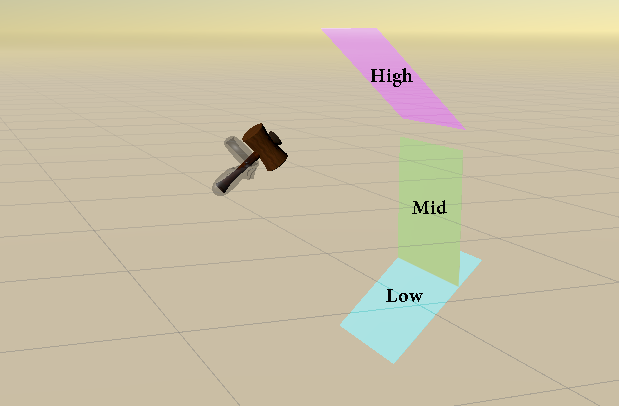}\label{fig:whac-a-mole-target-area-placement}}
	\caption{(a) In Whac-A-Mole, targets appear randomly at one of nine fixed positions and must be hit with a hammer within one second to score a point. The transparent targets are shown for visualisation purposes only and are not visible during the game. (b) The game allows for three different placements of the target area.}
	\Description{Top (subcaption: "Front View of Target Grid"): Front view of the target plane and the target grid in the Whac-A-Mole game, together with the controlled hammer hitting a target. Bottom (subcaption: "Target Area Placements"): Three target areas (low, mid, high) differing in terms of position and orientation are shown in 3D space together with the controlled hammer.}
\end{figure}

\subsection{Game Design}

In the Whac-A-Mole game the goal is to hit targets (the "moles") with a hammer to score points. The targets appear randomly on a $3\times3$ grid as shown in Figure~\ref{fig:whac-a-mole-front}. If a target is successfully hit within one second, it explodes and disappears (see Figure~\ref{fig:simulated-interaction-loop}) and the game score is increased by one. Otherwise, the target collapses and no points are given. One round of game play lasts for one minute.

We explore three design choices. First, the game has three difficulty levels defining the maximum number of simultaneously displayed targets (\textit{easy}: 1, \textit{medium}: 3, \textit{hard}: 5). %
The second design choice, target area placement, determines the position and orientation of the target area with respect to the HMD (\textit{low}, \textit{mid}, \textit{high}), as shown in Figure~\ref{fig:whac-a-mole-target-area-placement}. For the third design choice, we create two versions of the game. In the \textit{constrained} version, a velocity threshold needs to be exceeded in order to successfully hit a target, whereas in the \textit{unconstrained} version, no such threshold exists. Further details on the game design are provided %
in the supplementary material.

\subsection{Training Simulated Users}

We train a UitB user model to play the different variations of the game. We use \textit{MoblArmsWrist}, a MuJoCo version of the \textit{MoBL ARMS} model~\cite{Saul14} of the right arm, as the biomechanical model. This model includes shoulder, elbow and wrist joints, and the muscles actuating these joints, resulting in a total of 7 DOFs and 32 muscles. 
As the model does not actively model neck and head movements, the pose of the virtual HMD is constant throughout the simulation. \added{To account for this limitation, the target area was designed to be of a sufficiently small size to fit within the field of view of the participants, thus minimising the necessity for head movements. Furthermore, we use a non-scaled version of the biomechanical model, as we are not attempting to replicate a specific participant's movements, but rather aim to demonstrate \textsc{sim2vr}'s utility in generating movements that \textit{could be produced by a human participant}. Simulating the between-user variability by simultaneously training differently scaled biomechanical models is left for future work.}

We use the \texttt{UnityHeadset} vision module provided by \textsc{sim2vr}, along with the pre-implemented UitB proprioception module, to generate observations during training. The \texttt{UnityHeadset} module is configured to include the green colour and depth channels\footnote{\added{As the target colour changes from green to red, it is sufficient to use either the green or red channel to infer whether a target is about to disappear.}}, without stacking visual observations.
The proprioception module provides observations regarding the biomechanical model, such as the joint positions, velocities, and accelerations, as well as internal muscle activation states.

For training, we use a reward function consisting of multiple components, \added{which are obtained from states of the VR environment, such as game scores, and from states of the biomechanical model, such as muscle fatigue}:
\begin{equation}
r(s,a) = w_s S + w_c v_h C_c + w_d C_d + w_e C_e
\end{equation}
The component $S$ is the game score provided by the game; $C_c$ is a term to penalise target contacts with too low a hitting velocity in the \textit{constrained} version (to encourage faster hits); $C_d$ penalises the \added{sum of the }distance\added{s} between the hammer and each active target (to incentivise movement toward any target); and $C_e$ is an effort term based on the 3CC-r fatigue model~\cite{Looft18, Cheema20}.
The reward components are weighted with $w_s = 10$, $w_c = 2.5$, $w_d = 1$ and $w_e = 0.1$.
The target contact term $C_c$ is additionally weighted by the current hammer velocity $v_h$.

Simulated users were trained separately for the \textit{constrained} and \textit{unconstrained} versions, while keeping the difficulty level fixed to \textit{medium} and sampling the target area placements randomly at the beginning of each episode.\footnote{While it may be reasonable to continue training the user models on either the \textit{easy} or \textit{hard} difficulty levels, we observed that the models trained solely on the \textit{medium} level already generalise well to the remaining difficulties.}
For each version we train two instances of the simulated user, called SIM$_u$ and SIM$_a$. SIM$_u$ is trained on the default game dynamics, i.e., the position of the "mole" is sampled uniformly randomly among the nine possible positions. SIM$_a$, on the other hand, is trained using an adaptive automated curriculum manually implemented in Unity. Such a curriculum can be thought of as a ``personal trainer'' that creates a customised training plan for the simulated user based on \deleted{its}\added{their} current strengths and weaknesses. %

\begin{figure*}[h!]
	\centering
	\includegraphics[width=0.875\textwidth]{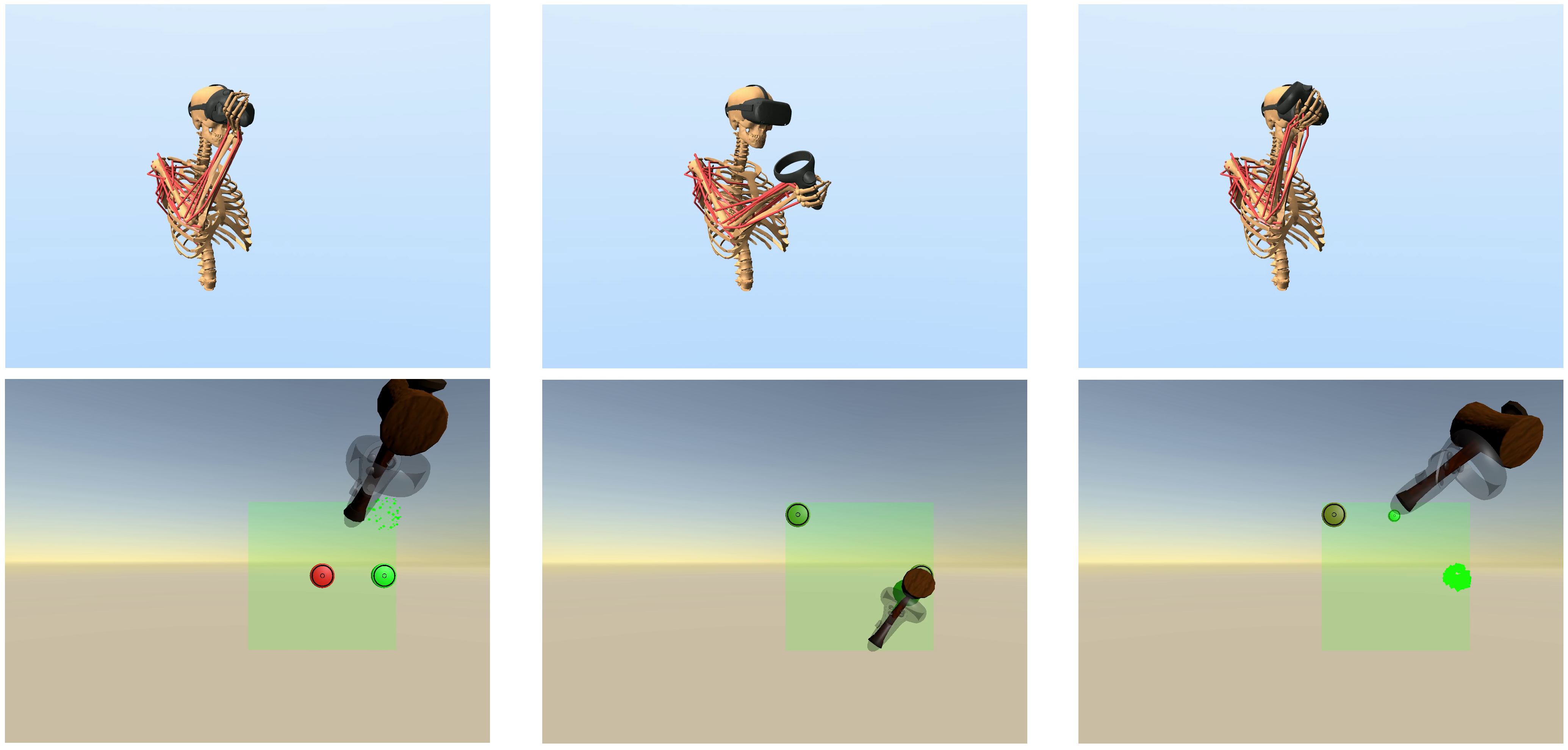}
	\caption{Still frames from an evaluation of a simulated user \deleted{as \deleted{it is}\added{they are} hitting a target on the right side of the target area}\added{playing Whac-A-Mole}. The top row shows the simulated user in MuJoCo, while the bottom row is a video capture from the Unity application. \added{In a typical hitting movement, the simulated user gains momentum (left), strikes the target with the virtual hammer (mid) and moves the hammer away from the target area by bending the elbow (right).}}
	\Description{2x3 still frames of a biomechanical user model equipped with the right VR controller and HMD playing the Whac-A-Mole game. The simulated user (top) and the corresponding VR application view (bottom) are shown below each other for three different time frames.}
	\label{fig:frames}
\end{figure*}

The learned control policies were evaluated for 1) all three difficulty levels while keeping the target area placement fixed to \textit{mid}, and 2) for all target area placements while keeping the difficulty level fixed to \textit{medium} (six configurations in total). %
Further details on the training and evaluation procedure can be found %
in the supplementary material.
A frame sequence of the simulated user playing the game with the learned control policy is shown in Figure~\ref{fig:frames}.

\subsection{User Study Method}\label{sec:whac-a-mole-user-study}

\deleted{In order to compare the simulated user behaviour against real player behaviour, we conducted a user study where we collected game playing data from 18 participants. 
Twelve participants played the \textit{constrained} version of the game, while six participants played the \textit{unconstrained} version. Each participant played the same six \textit{difficulty} $\times$ \textit{target area placement} configurations used to evaluate the simulated users.} %

\subsubsection*{Participants} \deleted{The participants were mostly local students (graduate or post-graduate) and faculty members at Aalto University.}
\added{In order to compare the simulated user behaviour against real player behaviour, we conducted a user study where we collected game playing data from 18 participants. All but two participants were students and faculty members at Aalto University; one participant was a student at the University of Helsinki, and one participant was not affiliated with either university.}
\added{All participants were right-handed.}
The average age was 28.8 years, with a standard deviation of 6.2 years. Eight of the participants identified as female, nine as male, and one as non-binary.

\subsubsection*{Experimental Design} \deleted{The participants first played the three difficulty levels, while keeping the target area placement fixed to \textit{mid}, in a counterbalanced order. 
Participants were then given a rest period of at least 5 minutes, with the option to rest longer if necessary to ensure their right arm was not fatigued. %
After resting, participants played the game in all three target area placements, in a counterbalanced order, with the difficulty level fixed to \textit{medium}, and reported the Borg Rating of Perceived Exertion (Borg RPE)~\cite{Borg82} score after each round.}
\added{Twelve participants played the \textit{constrained} version of the game, while six participants played the \textit{unconstrained} version.
Each participant played the six \textit{difficulty} $\times$ \textit{target area placement} configurations that were also used to evaluate the simulated users.
For each trial, we recorded the participant's body and hammer movement. For the three target area placement configurations (with \textit{medium} difficulty), we also asked participants to report the Borg Rating of Perceived Exertion (Borg RPE)~\cite{Borg82} score after each round.}

\subsubsection*{Procedure} The experiments were conducted in an office room. The participants were seated, but not restricted to the chair. Participants were asked to try to avoid bending and rotating their torso (as the simulated users have a fixed torso), and focus on their arm movements. Participants were allowed to train with the \textit{medium} difficulty level with \textit{mid} target area placement. This was to reduce learning effects, and to provide the participants a chance to familiarise themselves with the notion of hitting non-physical floating targets.
\added{The participants first played the three difficulty configurations in a counterbalanced order, and were then given a rest period of at least 5 minutes, with the option to rest longer if necessary to ensure their right arm was not fatigued.
Finally, participants played the three target area placement configurations in a counterbalanced order, and reported the Borg RPE after each round.
}
The experiment length was approximately 30 minutes, and participants were rewarded with a 15 euro gift card to a local restaurant.

\subsection{Results}\label{sec:whac-a-mole-evaluation}
We demonstrate the predictive capabilities of \textsc{sim2vr} by comparing the performance and effort estimates of simulated and real users in the different variations of the game.
We also %
show that \textsc{sim2vr} is capable of predicting specific strategies that users may exploit when game dynamics allow.

Unless otherwise stated, the following results were obtained with the \textit{constrained} game version.

\subsubsection{Performance}

\begin{figure}
	\centering
	\includegraphics[width=0.75\textwidth]{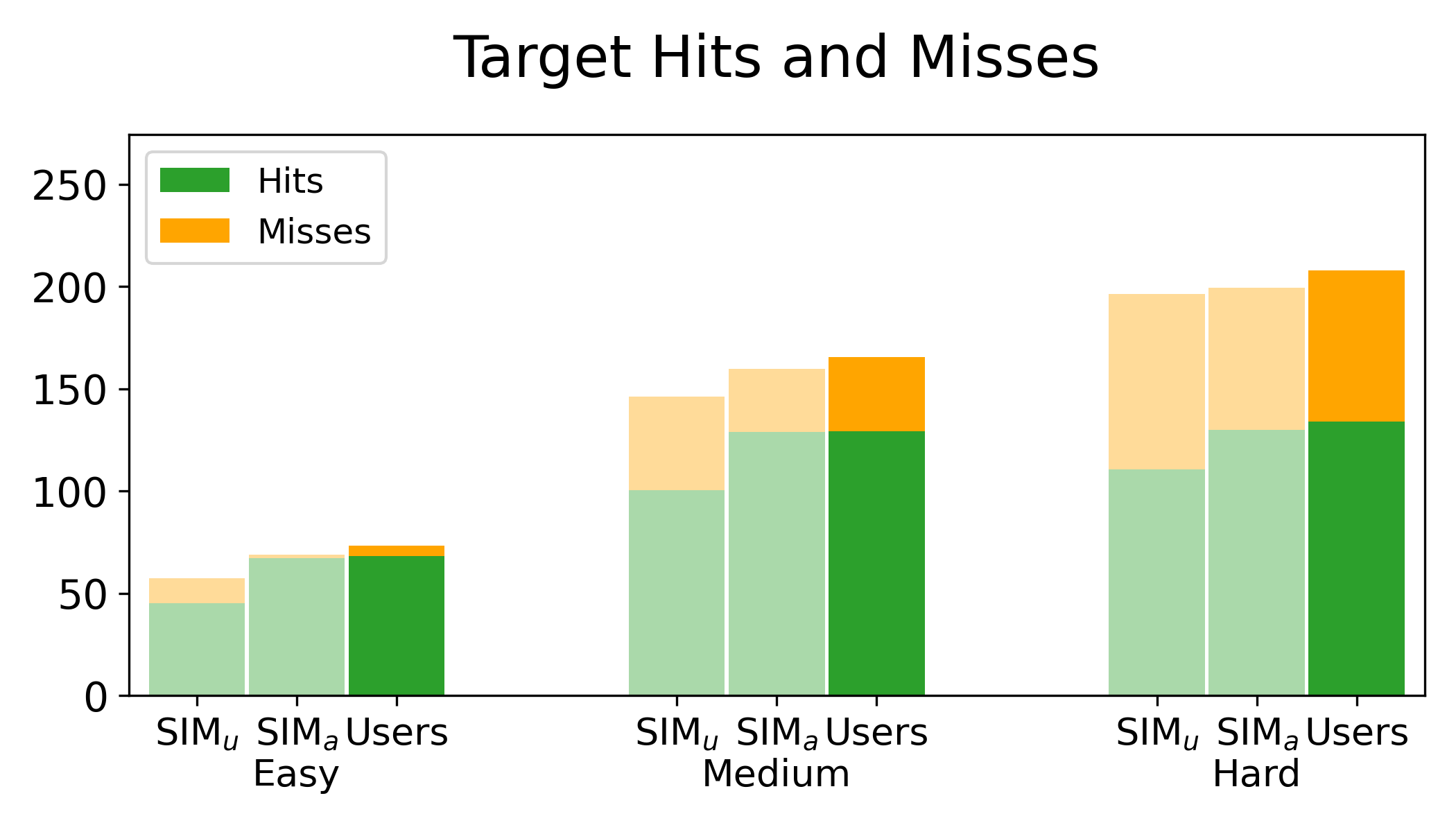}
	\caption{The simulated users SIM$_u$ and SIM$_a$ (left and mid bars) show similar numbers of target hits and misses in the \added{constrained version of the }Whac-A-Mole game as the 12 real users from our study on average (right bars). Differences between the three difficulty levels are predicted well.}
	\Description{Bar plot showing the average number of target hits and misses (y-axis; hits shown as green bars, misses shown as orange bars stacked on top) per difficulty level (x-axis; "Easy" on the left, "Medium" in the middle, "Hard" on the right) and for the two simulated users "SIM_u" and "SIM_a" and averaged across all real users (denoted as "Users"). Figure title: "Target Hits and Misses"}
	\label{fig:diff_sumstats}
\end{figure}

\begin{figure}
	\centering
	\includegraphics[width=0.75\textwidth]{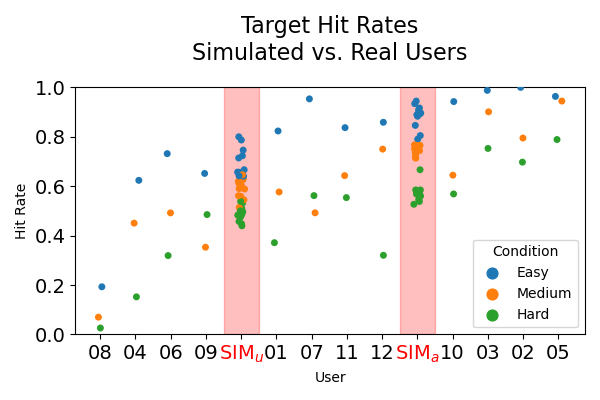}
	
	\caption{Both \added{simulated users }SIM$_u$ and SIM$_a$ \added{(highlighted in red) }predict target hit rates that are within the between-user variability observed from real players\added{ for the constrained version of the Whac-A-Mole game}. The simulations also show a similar decrease in hit rate as the game difficulty increases as most real players.
	}
	\Description{Scatter plot separately showing the target hit rate (y-axis) per difficulty level ("Easy" shown in blue, "Medium" shown in orange, "Hard" shown in green) for the 12 real users and the two simulated users "SIM_u" and "SIM_a", which are highlighted by vertical red stripes (x-axis). The 14 users are sorted in ascending order by the average hit rate across the three levels (08, 04, 06, 09, SIM_u, 01, 07, 11, 12, SIM_a, 10, 03, 02, 05). Figure title: "Target Hit Rates -- Simulated vs. Real Users"}
	\label{fig:diff_hitrates}
\end{figure}

In Whac-A-Mole, the performance (and thus the difficulty) of a game variant can be assessed by the number of target hits and misses users achieve per round.
In Figure~\ref{fig:diff_sumstats}, the mean number of hits and misses from 12 rounds of game play is shown for the two simulated users SIM$_u$ and SIM$_a$ for each level, along with the means of the 12 real users. For real players (solid, right bars), both target hits and misses increase from the \textit{easy} to the \textit{medium} level. However, this trend does not continue when proceeding to the \textit{hard} level. Instead, the number of hits remains approximately constant (green), while the number of misses (orange) increases. A similar trend can be observed for both SIM$_u$ and SIM$_a$ (shown as left and mid bars, respectively). Thus, if used early in the design process, the simulated users could have suggested that increasing the task difficulty beyond the \textit{medium} level would not lead to higher game scores for average users, but only to more target misses. %

In Figure~\ref{fig:diff_hitrates}, the target hit rate is plotted separately for all users in each difficulty level. While there is a considerable variability in the performance of real users, both simulations consistently predict hit rates that lie within this range. 
In addition, the decrease in hit rate with increasing game difficulty predicted by both simulated users is also observed for almost all real users (the only exceptions are users 9 and 11, who achieved higher hit rates on the hard than on the medium level). %

\subsubsection{Effort}

\begin{table*}
    \centering
    \begin{adjustbox}{max width=\textwidth}
    \begin{tabular}{c|c||c|p{50pt}|p{50pt}||p{50pt}|p{50pt}|p{50pt}}
         \multirow{2}*{\textbf{User}} & \multirow{2}*{\textbf{Variable}} & \multirow{2}{50pt}{\centering\textbf{Placement}} & \multirow{2}{50pt}{\centering\textbf{Mean/}\aptLtoX[graphic=no,type=html]{}{\\} \centering \textbf{Median$^\dagger$}} & \multirow{2}{50pt}{\centering\textbf{Std./}\aptLtoX[graphic=no,type=html]{}{\\}\textbf{IQR$^\dagger$}} & \multirow{2}{50pt}{\centering\textbf{Alternative} \aptLtoX[graphic=no,type=html]{}{\\} \centering \textbf{Hypothesis}} & \multicolumn{2}{c}{\textbf{Wilcoxon Signed Rank}} \\
         & & & & & 
         &\centering \textbf{Z-score} &\centering \textbf{p-value}\cr
         \Xhline{2\arrayrulewidth} %
         \multirow{3}*{SIM$_u$} & 
         \multirow{3}*{Max. Fatig. MUs} & Low & 0.232 & 0.001 & 
         Low < Mid & 3.059 & 1.0 (n.s.)\\ %
         & & Mid & 0.204 & 0.002 &
         Mid < High & -3.059 & \cellcolor{green!25} 0.0002 (***)\\ %
         & & High & 0.238 & 0.002 &
         Low < High & -3.059 & \cellcolor{green!25} 0.0002 (***)\\ %
         \hline
         \multirow{3}*{SIM$_a$}& 
         \multirow{3}*{Max. Fatig. MUs} & Low & 0.266 & 0.002 &
         Low < Mid & -2.51 & \cellcolor{green!25} 0.0046 (**)\\ %
         & & Mid & 0.269 & 0.002 &
         Mid < High & 3.059 & 1.0 (n.s.)\\ %
         & & High & 0.230 & 0.002 & 
         Low < High & 3.059 & 1.0 (n.s.)\\ %
         \hline
         \multirow{3}{40pt}{\centering User\aptLtoX[graphic=no,type=html]{}{\\} Study} &  
         \multirow{3}*{Borg RPE} & Low & 9$^\dagger$ & 1.25$^\dagger$ &
         Low < Mid & -1.508 & 0.0658 (n.s.)\\ %
         & & Mid & 9$^\dagger$ & 3$^\dagger$ &
         Mid < High & -2.414 & \cellcolor{green!25} 0.0079 (**)\\ %
         & & High & 9.5$^\dagger$ & 4.25$^\dagger$ &
         Low < High & -2.213 & \cellcolor{green!25} 0.0134 (*)\\ %
    \end{tabular}
    \end{adjustbox}
    \caption{
    Overview of both descriptive and inferential statistics for effort predictions. Descriptive statistics include the mean and standard deviation for the maximum fatigued motor units per round (averaged over all muscles), as predicted by the simulation, and the median and interquartile range (IQR) for the Borg RPE as reported in the user study ($^\dagger$)). Kolmogorov-Smirnov tests showed that none of the considered variables are normally distributed (all p-values < 0.001); we thus used one-sided Wilcoxon Signed rank tests to infer significant differences between the three target area placements. \\
    Explanations: *: $p\leq0.05$;
    **: $p\leq0.01$; ***: $p\leq0.001$; %
    n.s.: $p>0.05$. \\
    \textbf{Abbreviations:} \textit{Std.} for standard deviation, \textit{IQR} for \textit{interquartile range}, \textit{Max. Fatig. MUs} for \textit{maximum fatigued motor units} and \textit{n.s.} for \textit{not significant}.
    }\label{tab:whac-a-mole-effort}
    \Description{
The table consists of three parts. The left part has the columns "User" and "Variable" and the following rows:
    1. SIM_u - Max. Fatig. MUs.
    2. SIM_a - Max. Fatig. MUs.
    3. User Study - Borg RPE.
The second part, which is placed right next to the first part, has the columns "Placement", "Mean/Median †" and "Std./IQR †" and the following rows (three subrows per row of the first part):
    1a. Low - 0.232 - 0.001.
    1b. Mid - 0.204 - 0.002.
    1c. High - 0.238 - 0.002.
    2a. Low - 0.266 - 0.002.
    2b. Mid - 0.269 - 0.002.
    2c. High - 0.230 - 0.002.
    3a. Low - 9 † - 1.25 †.
    3b. Mid - 9 † - 3 †.
    3c. High - 9.5 † - 4.25 †.
The third part, which is placed right next to the first second, has the columns "Alternative Hypothesis", "Wilcoxon Signed Rank -- Z-score" and "Wilcoxon Signed Rank -- p-value" and the following rows (three subrows per row of the first part):
    1a. Low < Mid - 3.059 - 1.0 (n.s.).
    1b. Mid < High - -3.059 - 0.0002 (***).
    1c. Low < High - -3.059 - 0.0002 (***).
    2a. Low < Mid - -2.51 - 0.0046 (**).
    2b. Mid < High - 3.059 - 1.0 (n.s.).
    2c. Low < High - 3.059 - 1.0 (n.s.).
    3a. Low < Mid - -1.508 - 0.0658 (n.s.).
    3b. Mid < High - -2.414 - 0.0079 (**).
    3c. Low < High - -2.213 - 0.0134 (*).
In the "Wilcoxon Signed Rank -- p-value" column, all cells with at least one "*" (i.e., those showing significant differences) are highlighted in green.
    }
\end{table*}

To assess the impact of target area placement on movement effort, participants in the user study were asked to report their perceived exertion in terms of the Borg RPE scale, after playing each 1-minute round. The Borg RPE reported for the \textit{high} placement was significantly higher than for both the \textit{low} and \textit{mid} placements, while no significant difference could be found between \textit{low} and \textit{mid} placements (summary statistics as well as details on the statistical testing are given in Table~\ref{tab:whac-a-mole-effort}). This also agrees well with qualitative remarks of most participants, stating that the arm felt considerably more fatigued after playing the \textit{high} placement as compared to the \textit{low} and \textit{mid} placements. To estimate the level of exertion from our simulation, we measured the percentage of fatigued motor units as predicted by the 3CC-r fatigue model for each simulation time frame, calculated the maximum value per round for each muscle separately, and then took the average over all 32 muscles.

As shown in Table~\ref{tab:whac-a-mole-effort}, SIM$_u$ predicts a significant increase in fatigue between the \textit{mid} and \textit{high} as well as between the \textit{low} and \textit{high} placements, showing the same trend as observed for the Borg RPE scale for real players. Interestingly, the fatigue predicted for the \textit{low} placement is considerably higher than in \textit{mid} placement. Comparing the simulation videos for these placements, we found that in the \textit{low} placement SIM$_u$ follows a strategy that hardly flexes the elbow, but rather keeps the arm extended for almost the entire movement, which may lead to higher muscle fatigue over time. %

For SIM$_a$, the \textit{mid} placement exhibits significantly higher fatigue values than the \textit{low} placement ($p=0.0046$), whereas no significant increase from \textit{mid} to \textit{high} or from \textit{low} to \textit{high} could be found. We found that this is mainly due to different strategies for each of the three placements, despite being trained together and sharing the same network weights. In particular, SIM$_a$ has learned to efficiently reduce unnecessary shoulder movements mainly in the \textit{high} placement, resulting in a lower number of fatigued motor units than in the \textit{low} and \textit{mid} placements. 

Our findings can be seen as an example of how small changes in the training may have a considerable impact on which strategies are learned and predicted by the user simulation. While we demonstrate that characteristic differences in subjective levels of muscle exertion can be predicted in principle by our simulation-based approach, effort-related predictions must currently be treated with caution, as the effort can vary with small changes in the movement strategy.

\subsubsection{Strategy}

\begin{figure*}
	\centering
	\includegraphics[width=0.95\textwidth]{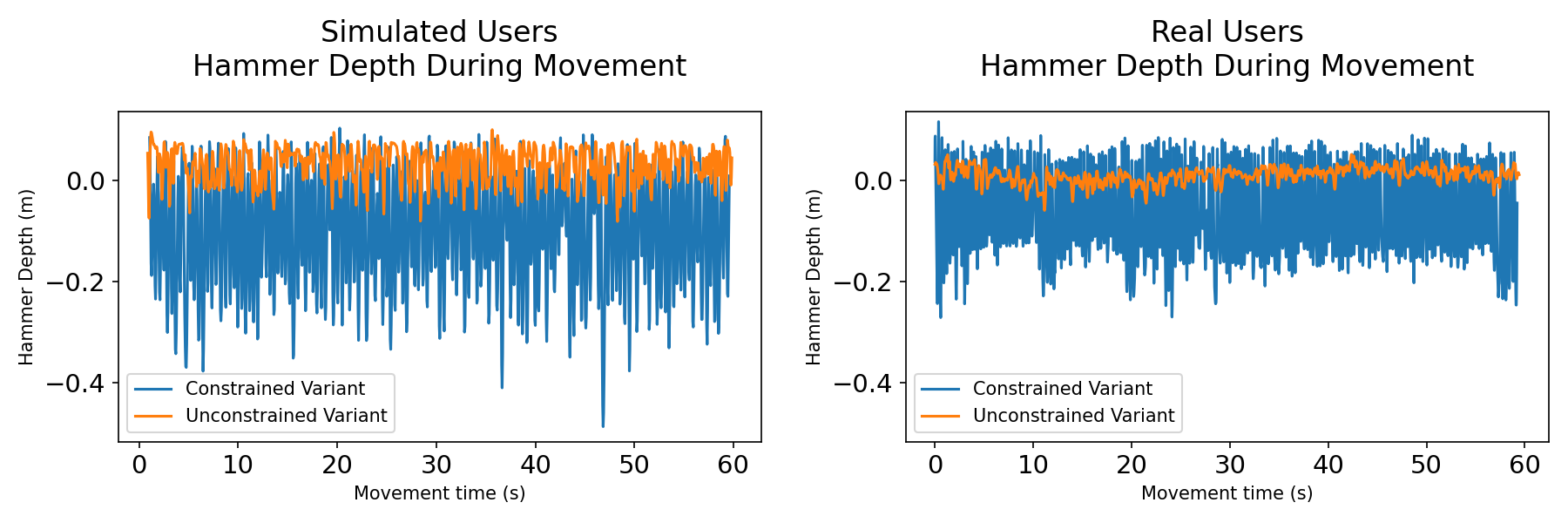}
	
	\caption{\added{Simulated users trained either on the constrained or unconstrained version of the Whac-A-Mole game predict different strategies. While the hammer is moved back and forth in the constrained version (blue line in left plot), it is held close to the target plane during the entire movement in the unconstrained version (orange line in left plot). This difference in strategy was also observed when comparing the movement of users playing the constrained and unconstrained versions (right plot). While both plots show %
			hammer movements for the \textit{hard} level, similar results were observed for the \textit{medium} level.}
	}
	\Description{
		Left (Subfigure title: "Simulated Users -- Hammer Depth During Movement"): Hammer depth in meters (y-axis) plotted over time (x-axis) for an entire round of 60 seconds, once for a simulated user movement in the constrained variant (blue line) and once in the unconstrained variant (orange line).
		Right (Subfigure title: "Real Users -- Hammer Depth During Movement"): Hammer depth in meters (y-axis) plotted over time (x-axis) for an entire round of 60 seconds, once for a real user movement in the constrained variant (blue line) and once in the unconstrained variant (orange line).}
	\label{fig:depth_strategy}
\end{figure*}

Simulations trained via \textsc{sim2vr} can anticipate the effect of a simple game design choice on the strategies employed by users. In Figure~\ref{fig:depth_strategy} this is shown for a trajectory comparison between the \textit{constrained} and \textit{unconstrained} versions of the game. In the \textit{unconstrained} version, an entirely different strategy is predicted by our simulation; in particular, the \textit{unconstrained} policy retracts the arm considerably less, and instead keeps the hammer relatively close to the target area during movements, which indeed allows to hit many more targets in the same time.\footnote{Applying the two simulation strategies to the \textit{hard} level, 195 target hits were observed in the \textit{unconstrained} variant, and 102 target hits and 35 additional target contacts with too low hitting velocity were observed in the \textit{constrained} variant.} %
Interestingly, one of the six users who played the \textit{unconstrained} version of the game exploited the same strategy, as shown in Figure~\ref{fig:depth_strategy} (right, orange line) along with a reference user trajectory from the \textit{constrained} version of the game (blue line).\footnote{Another user pursued a similar strategy but hit the targets more from above, while the remaining four users followed essentially the same strategy as the users in the \textit{constrained} version.}

This demonstrates the value of biomechanical user models in discovering specific strategies that users may exploit for a given game design. By making such insights available early in the VR development process, \textsc{sim2vr} can help find game dynamics that meet the designers' requirements for performance, effort, comfort, space limitations and more. %

\subsection{Summary}
This demonstration %
shows that \textsc{sim2vr} can be utilised to predict differences in terms of performance, effort and strategies between different variations of a VR game. Our approach enables the evaluation and comparison of different prototypes and design choices, and thus could be used in addition to (or as a partial replacement for) user studies in the early stages of application development.

\section{Discussion}\label{sec:discussion-limitations}

Computational approaches to modelling and simulation of users have the potential to be of value in design and analysis of interactive systems. However, in addition to the challenges of modelling the complexity of human behaviour, a significant challenge is modelling the sensory experience of users, who can be in complex environments and subject to non-modelled external disturbances to all senses, which can reduce the potential benefits of a ``sim-to-real'' transfer (a term mainly used in the field of robotics~\cite{Dimitropoulos22}).
VR interaction\deleted{, however,} is a special case of human-computer interaction in that significant aspects of the user's perception (namely vision and audio) can be controlled by the VR application to a greater extent than other cases. In particular, the (currently) low number of different VR systems, the ``full-screen mode'' of most VR applications and the typically high level of immersion~\cite{Seinfeld20} reduces the impact of external  disturbances when assessing performance. VR interaction thus offers ideal conditions for creating ecologically valid simulations and performing automated user testing. This makes VR a good starting point for research into simulation models which have well-aligned information from visual and auditory displays.

With \textsc{sim2vr}, we have built on these ideal conditions and developed a system that does not only give a biomechanical user model access to the same VR application as real users, but also ensures that it can interact with the application just as if a physical VR controller and HMD were used. %
This constitutes a decisive step towards automated biomechanical testing in VR.
In the future, biomechanical user simulations could lead to more comfortable and health-promoting VR interactions, as well as VR environments that specifically consider the abilities of individual user groups.

\subsection{Scope and Limitations}%

\added{\paragraph {Applying \textsc{sim2vr}}

It is important to note that the complexity of integrating \textsc{sim2vr} into an existing Unity application clearly depends on the application's structure. In some cases, additional scripts and methods might be required to properly initiate the application for the RL training process, including methods to skip main menus, select specific game levels, and reset to a specific initial state. This may also require modification of the source code, which is not always possible.

Furthermore, \textsc{sim2vr} does not currently support all forms of feedback and control, such as auditory and haptic feedback and usage of controller buttons. The integration of these into \textsc{sim2vr} is left for future work.
}
\deleted{Regarding the interaction technique, in this paper we have focused on VR interaction using standard VR controllers and HMD. An extension of \textsc{sim2vr} to other movement-based methods, such as full body movements sensed via Motion Capture, is straightforward.
The same applies to currently missing output modalities such as auditory or haptic feedback.}

\paragraph{\added{User Modelling}} %

\textsc{sim2vr} mainly addresses the application-related dimensions of the alignment problem\added{, reducing three potential error sources (unaligned input, differences in VR applications, unaligned output). This enables biomechanical forward simulations in a realistic VR interaction environment, as we have demonstrated in our Whac-A-Mole user study}. Aligning the user-related dimensions, i.e., physics, perception, and cognition, however, remains challenging. \added{This is mainly due to the immature state of accurate musculoskeletal and perception models, and incomplete knowledge of best practices in applying RL methods. In the following we will discuss how these limitations affect \textsc{sim2vr} and how they can be overcome.}\deleted{This particularly includes finding an appropriate biomechanical user model, modelling multisensory noisy perception, and designing a task-specific reward function, as we discuss in the following.}

Due to its dependency on UitB, \textsc{sim2vr} currently only provides the standard \textit{MoBL ARMS model}~\cite{Saul14}, which is unsuitable for simulating dexterous finger movements or body rotations. 
However, more sophisticated models of the human body focusing on either the elbow, hand, single finger or lower extremity have recently been released as part of the MyoSuite project\footnote{\url{https://sites.google.com/view/myosuite}} and can be integrated directly into UitB. 
In addition, the MyoConverter~\cite{Ikkala20,wang2022myosim} can be used to transfer OpenSim models to MuJoCo.

The vision module provided by \textsc{sim2vr} currently only models the user's visual perception of the virtual environment at a very rudimentary level.
To better reflect the complexity and subtlety of visuomotor VR interaction, this module should be extended in the future to include, for example, binocular, foveal, noisy and/or delayed visual observations.
Initial ideas for modelling these well-established phenomena might be derived from~\cite{Strasburger11, Kowler11, Fischer22}.

Simulating user cognition is known to be particularly challenging.
\deleted{In the context of reinforcement learning, a major part of cognition modelling is to identify appropriate reward terms that incentivise ``human-like'' behaviour.
However, no general guidelines have yet been proposed on how to define these reward terms appropriately for a given interaction task.}%
\added{In reinforcement learning, a major challenge in modelling cognition, beyond modelling internal representations, is the identification of appropriate reward terms.
However, no general guidelines have yet been proposed on how to define these rewards for a given interaction task.}
Additionally, reward functions can easily induce unintended behaviour, if not designed carefully.
For example, a reward signal that penalises the remaining distance to multiple positions at the same time (e.g., to all currently active targets) will encourage an "intermediate" strategy that opts for a particular target position later rather than sooner. %
Such a ``strategic bias'' may also degrade the quality of simulation-based predictions.  %

In order to expand the scope of automated biomechanical testing beyond simple visuomotor interaction tasks, further research into (and validation of) more complex RL-based training methods and design choices of reward functions and meta parameters is certainly needed.
Promising RL methods and frameworks in this regard include \textit{Hierarchical RL}~\cite{Li19, Nachum18}, in which a complex learning problem is decomposed into simpler subtasks that are easier to learn, and \textit{Transfer Learning}~\cite{Zhu20}, in which models are first exposed to a data-rich domain before being fine-tuned in a data-scarce or otherwise more challenging environment.
Moreover, recent successes in training user models for cognitively and biomechanically challenging motor control and interaction tasks, such as rotating multiple Baoding balls over the palm of the hand~\cite{MyoChallenge2022} or parking a remote-controlled car using a joystick~\cite{Ikkala22}, raise hopes for versatile simulated users that can be integrated directly into a range of existing VR applications without extensive refinement.

\deleted{The availability of highly skilled and approved simulated users for a variety of interaction tasks will \deleted{therefore }be critical to the extent to which biomechanical simulation finds its way into future automated VR testing routines.}
\added{These examples demonstrate a growing community effort, from an increasing range of disciplines. Future community effort towards highly skilled and approved simulated users for a variety of interaction tasks will determine the extent to which biomechanical simulation finds its way into future automated VR testing routines.}

\section{Conclusion}
\label{sec:conclusion}
\added{While simulating dynamic interactive behaviour on a visuomotor level is a hard problem that will take years to solve in its entirety, biomechanical user simulations provide enormous potential both for researchers and interaction designers.} %
The \textsc{sim2vr} system presented in this paper \added{constitutes a key first step towards automated biomechanical testing, as it }enables the training of RL-based simulated users directly in the same VR environments that real users interact with.

\deleted{By partially addressing the VR Simulation Alignment Problem,}\added{By addressing the application-related dimensions of the alignment problem,} our approach increases the ecological validity of biomechanical user simulations in VR. %
Using Whac-A-Mole as an example, we have demonstrated that \textsc{sim2vr} can predict characteristic differences between different game variants in terms of performance and effort, and reveal potential user strategies.

By lowering the barrier to the use of biomechanical simulation, we hope that \textsc{sim2vr} will \added{incentivise more research into better theories and models of user interaction and }increase the adoption of automated testing methods in VR development.
Given that biomechanical user models can help identify bugs, safety risks and health issues prior to user testing, we expect user modelling to assume a position as an indispensable part of the VR development process in the future.

\begin{acks} 
A.I. received funding from the Academy of Finland Flagship programme ``Finnish Center for Artifcial Intelligence'' (FCAI). 
P.H. has been supported by the European Commission through the Horizon 2020 FET Proactive program (grant agreement 101017779).
R.M.S. and M.K. received funding from the DIFAI ERC Advanced Grant (proposal 101097708, funded by the UK Horizon guarantee scheme as EPSRC project EP/Y029178/1). R.M.S. also received funding from EPSRC projects EP/T00097X/1, EP/R018634/1, and EP/T021020/1.
A.O. was funded by Research Council of Finland (grant no. 357578) and ERC (grant no. 101141916).
\end{acks}


\end{document}


\title[SIM2VR-SuppMat]{Supplementary Material:\\SIM2VR: Towards Automated Biomechanical Testing in VR}

\author{Florian Fischer}
\authornote{Both authors contributed equally to this research.}
\orcid{0000-0001-7530-6838}
\affiliation{%
 \institution{University of Cambridge}
 \country{United Kingdom}
 }

\author{Aleksi Ikkala}
\authornotemark[1]
\orcid{0000-0001-9118-1860}
\affiliation{%
  \institution{Aalto University}
  \country{Finland}
  }

\author{Markus Klar}
\orcid{0000-0003-2445-152X}
\affiliation{%
 \institution{University of Glasgow}
 \country{Scotland}
 }

\author{Arthur Fleig}
\orcid{0000-0003-4987-7308}
\affiliation{%
 \institution{ScaDS.AI Dresden/Leipzig} %
 \country{Germany}
 }
 
\author{Miroslav Bachinski}
\orcid{0000-0002-2245-3700}
\affiliation{%
 \institution{University of Bergen}
 \country{Norway}
 }
 
\author{Roderick Murray-Smith}
\orcid{0000-0003-4228-7962}
\affiliation{%
  \institution{University of Glasgow}
  \country{Scotland}
  }

\author{Perttu H\"{a}m\"{a}l\"{a}inen}
\orcid{0000-0001-7764-3459}
\affiliation{%
  \institution{Aalto University}
  \country{Finland}
  }

\author{Antti Oulasvirta}
\orcid{0000-0002-2498-7837}
\affiliation{%
  \institution{Aalto University}
  \country{Finland}
  }

\author{J\"{o}rg M\"{u}ller}
\orcid{0000-0002-4971-9126}
\affiliation{%
 \institution{University of Bayreuth}
 \country{Germany}
 }

\renewcommand{\shortauthors}{Fischer and Ikkala, et al.}

\maketitle

\section{Creating the Simulated User: Additional Tools}
\subsection{Reach Envelope Tool}
\begin{figure}[h!]
    \centering
    \includegraphics[width=0.65\textwidth]{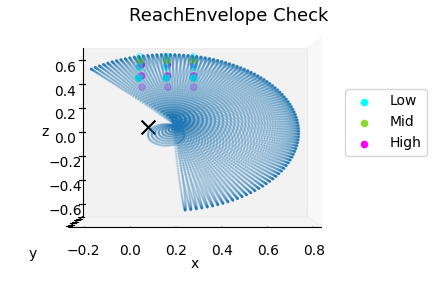}
    \caption{The reach envelope tool provided by \textsc{sim2vr} visualises the range that the simulated user (black cross, facing in z-axis direction) can theoretically reach with a maximally extended arm (blue lines) along with target positions relevant for the considered VR application (coloured dots). This example shows that in the Whac-A-Mole game from the main paper, all target positions in all three target area placements can be reached by the simulated user, albeit some of the low targets (those close to the boundary of the envelope) require a fully extended arm.
    }
    \Description{In a 3D plot, the area that can be reached by the simulated user with maximally extended arm is shown as blue lines from a bird's eye view. A black cross symbolises the position of the simulated user. The figure suggests that the target positions of all three target area placement conditions in the Whac-A-Mole game (visualised by cyan, green and magenta dots for low, mid and high conditions, respectively) are within this reach envelope. Figure title: "ReachEnvelope Check"}
    \label{fig:whac-a-mole-reach-envelope}
\end{figure}

As the biomechanical model typically has fewer degrees-of-freedom (DOFs) than a real human body, it is important to ensure that it is still able to perform all the movements required by the interaction. 
For a model of the upper extremity used to simulate visuomotor interaction in VR, this mainly involves verifying that all target positions are within the model's reach envelope.
The \textit{reach envelope tool} provided by \textsc{sim2vr} allows interactive exploration and visualisation of the body-centred positions that a given simulated user can theoretically reach. This can be used to verify that all target positions that occur in a given VR interaction task can be reached by the simulated user.
The tool works as follows.

Given a simulated user, \deleted{its}\added{the} reach envelope is computed by identifying all VR controller positions that can be reached with a maximally extended arm. 
These positions are then transformed into Unity coordinates and plotted in 3D along with desired target positions (see Figure~\ref{fig:whac-a-mole-reach-envelope} for an example plot; the tool can also create an animated 360° plot for better spatial perception).
Target positions can either be provided manually (in Unity coordinates) or can be directly extracted from a Unity log file.
The tool also allows one to distinguish between the envelope of the pure biomechanical model and the envelope when equipped with a VR controller, which can be used to verify correct placement of the controller mesh.

Running this tool for a given simulated user and VR interaction task can help improve the physical alignment. %

\subsection{Reward Scaling Tool}
Reward functions required to learn biomechanically plausible behaviour typically involve multiple components.
For example, effort costs are usually added to the reward to incentivise accomplishing the task with as little muscle effort as possible~\cite{Todorov02, Croxson09}.
However, balancing these reward terms is a non-trivial task and typically requires some RL experience. 

The provided \textit{reward scaling tool} may help in this regard. It takes the selected reward components (e.g., a one-time bonus reward given at task completion, a dense reward continuously awarded based on the current distance to all active targets, and the effort costs) as input, and then predicts how the contribution of each component to the total reward evolves over time. Predictions are generated by evaluating the reward components for different scenarios such as worst case (i.e., when the simulated user has not learned to make any movements yet), best case (i.e., when the desired target position is already reached at beginning of an episode), and linear or quadratic interpolation between initial and target position. The resulting plots can be used to identify poorly scaled reward components that either have too large of an effect on the simulated user or no effect at all, in which case these components need to be differently weighted. %

By helping to identify a reasonable trade-off between different reward components, this tool represents a first step towards better cognitive alignment between simulated and real users. %
A discussion on the current capabilities of simulated users for modelling VR interaction, and how these might benefit from more sophisticated RL methods, can be found in Section~6 in the main paper. %

\section{Details of the "Whac-A-Mole" Example}\label{sec:suppmat-whac-a-mole} %

\subsection{Game Design}\label{sec:whac-a-mole-game-design}

\subsubsection*{Targets}

Each target has a radius of 2.5cm and a life span of one second, during which it linearly changes its colour from green to red. Targets are randomly spawned at variable time intervals between 0 and 0.5 seconds, unless the maximum number of simultaneous targets (which depends on the selected \textit{difficulty} level) has already been reached. Also, if there are zero targets in the area (i.e., all targets have been hit), a new target will be spawned instantaneously.

\subsubsection*{Target Area Placement}

The target area placement defines the position and orientation of the target area with respect to the HMD. All areas are positioned 15cm to the right of the HMD, to ensure better reachability for the right arm. In the \textit{low} placement, the area is located 30cm below and 35cm in front of the HMD, tilted 45\textdegree; in the \textit{mid} placement, the area is located 10cm below and 40cm in front of the HMD, facing the user; and in the \textit{high} placement, the area is located 20cm above and 30cm in front of the HMD, tilted -45\textdegree. The different target area placements are depicted in Figure 6 (main paper). These poses were set based on feedback from pilot sessions.

\subsubsection*{Velocity Constraint}
We defined both a \textit{constrained} and an \textit{unconstrained} version of the game. In the \textit{constrained} version, a velocity threshold needs to be exceeded in order to successfully hit a target, while in the \textit{unconstrained} version all target contacts count as hits. In the \textit{constrained} version, targets need to be hit with a minimum speed of 0.8 \unit{\meter\per\second} in downward direction for the \textit{low} placement, and with a minimum speed of 0.8 \unit{\meter\per\second} in forward direction for the \textit{mid} and the \textit{high} placements. These values and directions were set based on feedback from pilot sessions.

\subsection{Training the Simulated User}

\subsubsection*{Adaptive Automated Curriculum}
For training the simulated user with the adaptive automated curriculum (SIM$_a$), target positions are sampled uniformly randomly with a probability of 50\%, and otherwise higher probabilities were assigned to those positions that had higher fail rates in the previous episode (i.e., a larger percentage of missed targets, relative to the percentage of missed targets at any target position). %

\subsubsection*{RL Training Hyperparameters}
All models were trained for 100M steps with default UitB training hyperparameters (see Table~\ref{tab:hyperparameters}) to ensure comparability. %

\begin{table}[!h]
\caption{UitB hyperparameters used to train the simulated users.}

\begin{adjustbox}{max width=\textwidth}
\begin{tabular}{p{4cm}p{8cm}}\label{tab:hyperparameters}
\textbf{Hyperparameter}           & \textbf{Value}

\\ \hline
Learning algorithm                & PPO

\\
Policy network architecture       & 256 x 256 (shared layers), preceded by a linear layer of size 128 for encoding observations from the proprioception module (\texttt{BasicWithEndEffectorPosition}), and a CNN of size (8, 16, 32) followed by a linear layer of size 256 for encoding observations from the vision module (\texttt{UnityHeadset}) 

\\
Visual observation size           & 120x80 pixels

\\
Timesteps                         & 100M                                                                       

\\
Learning rate schedule            & Linear, decay starts at 20M steps                                          

\\
Initial learning rate             & 5e-5                                                                       

\\
Final learning rate               & 1e-7                                                                       

\\
\# parallel environments   & 10

\\
\# steps per environment   & 4000                                                                              

\\
Batch size                        & 1000                                                                       

\\
KL divergence limit  & 1.0                                                                                                                                                                                     
\end{tabular}
\end{adjustbox}
\end{table}

\subsubsection*{Performance Measurements}\label{sec:performance-measurements}
Using a workstation with an AMD Ryzen Threadripper 3970X 32-core processor\footnote{\href{https://www.amd.com/}{https://www.amd.com/}}, and setting the number of parallel workers to 10, training a simulated model for 100M steps from scratch took approximately 48-72 hours and required 3.2GB memory on the GPU (NVIDIA GeForce RTX 3090\footnote{\href{https://www.nvidia.com/}{https://www.nvidia.com/}}). %
Most modern GPUs should thus be capable of several training runs in parallel, which, e.g., can be used to obtain a more robust measure of the quality of a given simulated user model by running the same training with different seeds.

Evaluating a learned policy took 30-40 seconds for a one-minute round, depending on whether a video was recorded simultaneously.

\subsection{Additional Results}\label{sec:suppmat-other-findings}
In addition to predicting differences in performance and effort and revealing potential user strategies, as shown in the main paper, our proposed approach can also be used to infer general characteristics of movement trajectories during interaction. %

\begin{figure*}[!h] %
    \centering
    \includegraphics[width=0.49\textwidth]{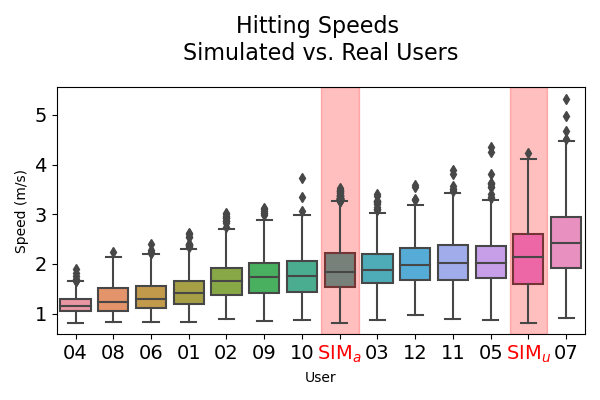}
    \includegraphics[width=0.49\textwidth]{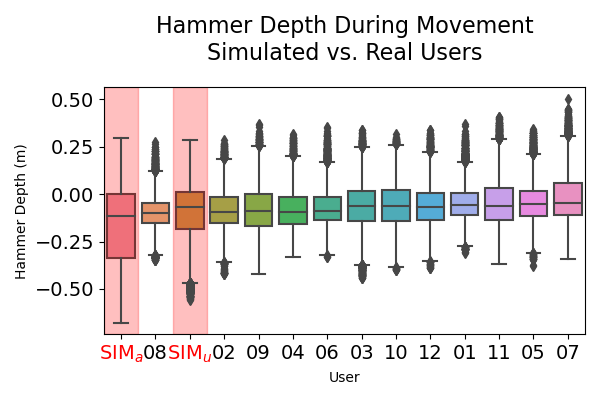}
    
    \caption{
    \textbf{Left:} The speed at which a target is (successfully) hit varies greatly from user to user. The hitting speed predicted by the simulation is within this between-user variability.
    \textbf{Right:} The hammer depth during the movement show comparable values between simulated and real users. This indicates that the simulation has learned to retract the arm to a similar extent as the users in our study.
    }
    \Description{
Left (Subfigure title: "Hitting Speeds -- Simulated vs. Real Users"): Box plot separately showing the speed in m/s (y-axis) for the 12 real users and the two simulated users "SIM_u" and "SIM_a", which are highlighted by vertical red stripes (x-axis). The 14 users are sorted in ascending order by the average speed (04, 08, 06, 01, 02, 09, 10, SIM_a, 03, 12, 11, 05, SIM_u, 07).
Right (Subfigure title: "Hammer Depth During Movement -- Simulated vs. Real Users"): Box plot separately showing the hammer depth in m (y-axis) for the 12 real users and the two simulated users "SIM_u" and "SIM_a", which are highlighted by vertical red stripes (x-axis). The 14 users are sorted in ascending order by the average hammer depth (SIM_a, 08, SIM_u, 02, 09, 04, 06, 03, 10, 12, 01, 11, 05, 07).}
    \label{fig:diff_hitspeeds_depthdistance}

    \centering

    \includegraphics[width=0.9\textwidth]{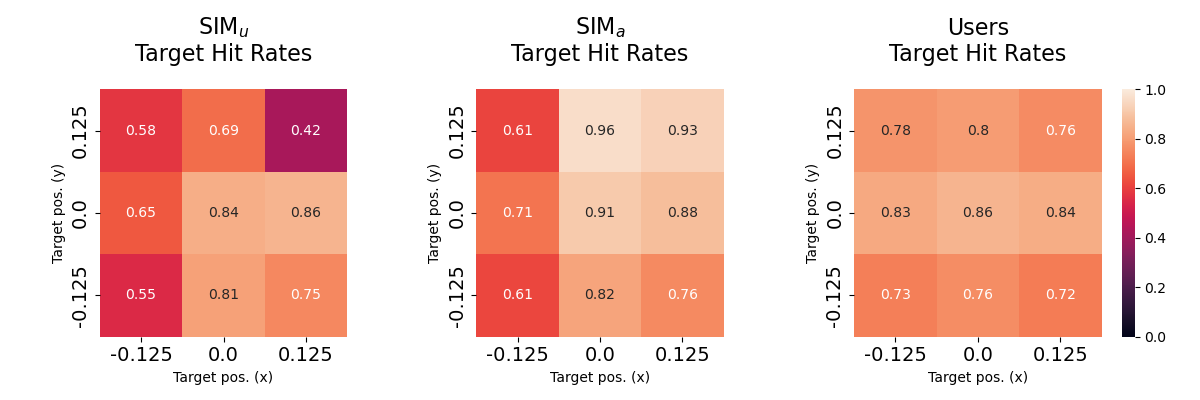}
    
    \caption{
    The hitting rates crucially differ depending on where targets are located relative to the shoulder. In the user study (right), low targets were hit less often than mid and high targets on average. This is captured well by the simulations (left and mid), which, however, also exhibit lower hitting rates for left targets. This could indicate areas that are more difficult for the biomechanical model to reach and/or control policies that did not converge to the (global) optimum.}
    \label{fig:diff_hitrates_directional}
    \Description{
Left (Subfigure title: "SIM_u -- Target Hit Rates"): 3x3 heatmap with target position in x-/y-dimension on x-/y-axis (tick labels for both axis: -0.125, 0, 0.125), a colour map in the range from 0 to 1 and the following values from top left to bottom right (row-wise): 0.58 - 0.69 - 0.42 -- 0.65 - 0.84 - 0.86 -- 0.55 - 0.81 - 0.75.
Middle (Subfigure title: "SIM_a -- Target Hit Rates"): 3x3 heatmap with target position in x-/y-dimension on x-/y-axis (tick labels for both axis: -0.125, 0, 0.125), a colour map in the range from 0 to 1 and the following values from top left to bottom right (row-wise): 0.61 - 0.96 - 0.93 -- 0.71 - 0.91 - 0.88 -- 0.61 - 0.82 - 0.76.
Right (Subfigure title: "Users -- Target Hit Rates"): 3x3 heatmap with target position in x-/y-dimension on x-/y-axis (tick labels for both axis: -0.125, 0, 0.125), a colour map in the range from 0 to 1 and the following values from top left to bottom right (row-wise): 0.78 - 0.8 - 0.76 -- 0.83 - 0.86 - 0.84 -- 0.73 - 0.76 - 0.72.}

    \centering
    \includegraphics[width=0.9\textwidth]{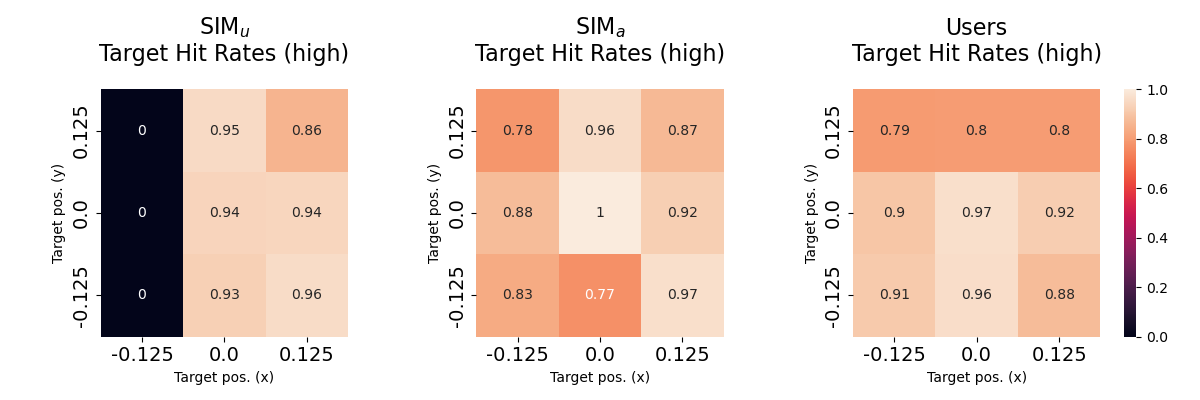}
    \caption{While SIM$_u$ does not reach the left targets in the high target area placement at all (left), SIM$_a$ shows a plausible average target hit rate for any target location (cf. mid and right plot).
    }
    \label{fig:diff_hitrates_directional_high}
    \Description{
Left (Subfigure title: "SIM_u -- Target Hit Rates (high)"): 3x3 heatmap with target position in x-/y-dimension on x-/y-axis (tick labels for both axis: -0.125, 0, 0.125), a colour map in the range from 0 to 1 and the following values from top left to bottom right (row-wise): 0 - 0.95 - 0.86 -- 0 - 0.94 - 0.94 -- 0 - 0.93 - 0.96.
Middle (Subfigure title: "SIM_a -- Target Hit Rates (high)"): 3x3 heatmap with target position in x-/y-dimension on x-/y-axis (tick labels for both axis: -0.125, 0, 0.125), a colour map in the range from 0 to 1 and the following values from top left to bottom right (row-wise): 0.78 - 0.96 - 0.87 -- 0.88 - 1 - 0.92 -- 0.83 - 0.77 - 0.97.
Right (Subfigure title: "Users -- Target Hit Rates (high)"): 3x3 heatmap with target position in x-/y-dimension on x-/y-axis (tick labels for both axis: -0.125, 0, 0.125), a colour map in the range from 0 to 1 and the following values from top left to bottom right (row-wise): 0.79 - 0.8 - 0.8 -- 0.9 - 0.97 - 0.92 -- 0.91 - 0.96 - 0.88.}
\end{figure*}

In Figure~\ref{fig:diff_hitspeeds_depthdistance} (left), the speed of the hammer when hitting a target is shown for all simulated and real users. The hitting speeds predicted by SIM$_u$ and SIM$_a$ are within the range of the real users, both in terms of mean and variance. While the predictions of SIM$_u$ are at the upper end of this range, those of SIM$_a$ are closer to the average hitting speed.
This shows that both simulations generate reasonable hitting speeds rather than exhibiting movements beyond (or below) human capabilities.

We also found the ``hammer depth'' -- the position of the hammer in the forward-backward direction, relative to the target plane -- to be a key factor in assessing the quality of a learned simulation. As shown in Figure~\ref{fig:diff_hitspeeds_depthdistance} (right), all 12 real users exhibit similar depth offsets in terms of range and mean when playing the game. The hammer depths of the two simulation models SIM$_u$ and SIM$_a$ also show comparable values, indicating that the simulations retract the arm in a similar way as real users.

An interesting behaviour to assess would be which targets simulated users and real players aim for. 
As shown in Figure~\ref{fig:diff_hitrates_directional} (right), the three lower targets exhibit a lower average hit rate than the three centred target positions for real users in %
the \textit{constrained} version.
This trend is also visible in the evaluation of SIM$_u$ (left) and SIM$_a$ (mid). In addition, both simulated users show a clear preference towards targets in the mid and right column, whereas targets on the left are more frequently ignored. This strategy was not observed in the user study, but in our experience may be partly due to 
the simulated user's general difficulty in reaching targets on the left-front half-sphere. %
In general, differences in target hit rates between specific target positions could be caused by policies getting stuck in local minima during training. In the high condition, for example, SIM$_u$ is not able to reach the left targets at all, while SIM$_a$ has learned to do so (cf. Figure~\ref{fig:diff_hitrates_directional_high}).
